\documentclass[sigconf]{acmart}

\usepackage{subcaption}
\usepackage{balance}
\usepackage{tabularx}
\usepackage[export]{adjustbox}
\usepackage{adjustbox}
\usepackage{multirow}
\usepackage[font={small,it},labelfont=bf]{caption}
\usepackage{dblfloatfix}
\usepackage{xcolor,colortbl}
\usepackage[flushleft]{threeparttable}

\newcommand{\seeddomains}{268}
\definecolor{Gray}{gray}{0.85}
\begin{document}

\copyrightyear{2017}
\acmYear{2017}
\setcopyright{acmcopyright}
\acmConference{CCS '17}{October 30-November 3, 2017}{Dallas, TX, USA}
\acmPrice{15.00}
\acmDOI{10.1145/3133956.3134002}
\acmISBN{978-1-4503-4946-8/17/10}

\fancyhead{}
\settopmatter{printacmref=false, printccs=true, printfolios=true} 

\title{Hiding in Plain Sight:\\A Longitudinal Study of Combosquatting Abuse}

\author{Panagiotis Kintis}
\affiliation{\institution{Georgia Institute of Technology}}
\email{kintis@gatech.edu}
\author{Najmeh Miramirkhani}
\affiliation{\institution{Stony Brook University}}
\email{nmiramirkhani@cs.stonybrook.edu}
\author{Charles Lever}
\affiliation{\institution{Georgia Institute of Technology}}
\email{chazlever@gatech.edu}
\author{Yizheng Chen}
\affiliation{\institution{Georgia Institute of Technology}}
\email{yzchen@gatech.edu}
\author{Roza Romero-G\'omez}
\affiliation{\institution{Georgia Institute of Technology}}
\email{rgomez30@gatech.edu}
\author{Nikolaos Pitropakis}
\affiliation{\institution{London South Bank University}}
\email{pitropan@lsbu.ac.uk}
\author{Nick Nikiforakis}
\affiliation{\institution{Stony Brook University}}
\email{nick@cs.stonybrook.edu}
\author{Manos Antonakakis}
\affiliation{\institution{Georgia Institute of Technology}}
\email{manos@gatech.edu}


\begin{abstract}
  Domain squatting is a common adversarial practice where attackers register
  domain names that are purposefully similar to popular domains. In this work,
  we study a specific type of domain squatting called ``combosquatting,'' in
  which attackers register domains that combine a popular trademark with one or
  more phrases (e.g., betterfacebook[.]com, youtube-live[.]com). We perform the
  first large-scale, empirical study of combosquatting by analyzing more than
  468 billion DNS records---collected from passive and active DNS data sources
  over almost six years. We find that almost $60\%$ of abusive combosquatting
  domains live for more than 1,000 days, and even worse, we observe increased
  activity associated with combosquatting year over year. Moreover, we show that
  combosquatting is used to perform a spectrum of different types of abuse
  including phishing, social engineering, affiliate abuse, trademark abuse, and
  even advanced persistent threats. Our results suggest that combosquatting is a
  real problem that requires increased scrutiny by the security community.
\end{abstract}

\maketitle

\section{Introduction}
\label{section:introduction}

The Domain Name System (DNS)~\cite{rfc882,rfc883}, is a distributed hierarchical
database that acts as the Internet's phone book. DNS's main goal is the
translation of human readable domains to IP addresses. The reliability and
agility that DNS offers has been fundamental to the effort to scale information
and business across the Internet. Thus, it is not surprising that miscreants
heavily rely on DNS to scale their abusive operations.

In fact, domain squatting is a very common tactic used to facilitate abuse by
registering domains that are confusingly similar~\cite{acpa} to those belonging
to large Internet brands. Past work has thoroughly investigated typosquatting
(domain squatting via typographical
errors)~\cite{agten2015seven,wang2006strider,edelman2003,khan2015every,Szurdi:long-taile-of-typosquatting,Moore2010},
bit squatting (domain squatting via accidental bit
flips)~\cite{bitsquatting_www2013,dinaburg:20011}, homograph-based squatting
(domains that abuse characters from different character
sets)~\cite{Holgers:2006:CTC:1267359.1267383,Gabrilovich:2002:HA:503124.503156},
and homophone-based squatting (domains that abuse the pronunciation similarity
of different words)~\cite{nikiforakis2014soundsquatting}.

A type of domain squatting that has yet to be extensively studied is that of
``combosquatting.'' Combosquatting refers to the combination of a recognizable
brand name with other keywords (e.g., paypal-members[.]com and
facebookfriends[.]com). While some existing research uses other terms to
describe combosquatting domains (i.e.,, ``cousin
domains''~\cite{jakobsson2007human}), this work only studies combosquatting in
the context of phishing abuse, failing to capture the full spectrum of potential
abuse. Thus, even though the general concept of constructing these types of
malicious domains is part of the collective consciousness of security
researchers, the community lacks a large-scale, empirical study on
combosquatting and how it may be abused. Therefore, the security community has
little insight into which trademarks domain squatters commonly abuse, how well
existing blacklists capture such abuse, and which types of abuse combosquatting
is used for.

In this work, we conduct the first large-scale, longitudinal study of
combosquatting abuse to empirically measure its impact. By combining more than
468 billion DNS records from both active and passive DNS datasets, which span
almost six years, we identify 2.7 million combosquatting domains that target
\seeddomains{} of the most popular trademarks in the US, and we find that
combosquatting domains are 100 times more prevalent than typosquatting
domains---despite the fact that combosquatting has been less studied. Our study
also makes several key observations that help better characterize how
combosquatting is used for abuse.

First, we study the \emph{lexical characteristics} of combosquatting domains. We
observe that combosquatting lacks generative models and find that, while
combosquatting domains vary in overall length, 50\% add at most eight additional
characters to the original trademark being abused. Furthermore, 40\% of
combosquatting domains are constructed by adding a single token
(Section~\ref{sec:lexical}) to the original trademark. Thus, while the pool of
potential combosquatting domains is very large, we find that many instances of
combosquatting try and limit the overall length of the combosquatting domain.
Additionally, we find that combosquatting domains tend to prefer words that are
closely related to the underlying business category of the trademark---resulting
in combinations that are more targeted than random.

Second, we analyze the \emph{temporal properties} of combosquatting domains and,
surprisingly, we see that almost 60\% of the abusive combosquatting domains can
be found in our datasets for more than 1,000 days---suggesting that these
abusive domains can often go unremediated. When combosquatting domains \emph{do}
become known to the security community, it is often significantly after the
threat was seen in the wild. For example, 20\% of the abusive combosquatting
domains appear on a public blacklist almost 100 days after we observe initial
resolutions in our DNS datasets, and this number goes up to 30\% for
combosquatting domains observed in malware feeds. To make matters worse, we
observe a growing number of queries to combosquatting domains year over year,
which is in stark contrast to better known squatting techniques like
typosquatting. Thus, combosquatting appears to be an increasingly effective
technique used by Internet miscreants.

Third, we discover and analyze numerous instances of combosquatting abuse in the
real world. Through a substantial crawling and manual labeling effort, we
discover that combosquatting domains are used to perform many different types of
abuse that include phishing, social engineering, affiliate abuse, and trademark
abuse (i.e., capitalizing on the popularity of trademarks to sell their own
products and services). By analyzing publicly available threat reports, we also
identified 65 combosquatting domains that were used by Advanced Persistent
Threat (APT) campaigns. These findings highlight the wide reaching impact of
combosquatting abuse. Finally, we manually analyzed various techniques attackers
used to drop malware and counter detection---leading to some interesting
discoveries surrounding the use of redirection chains and cookies.

In summary, combosquatting is a type of domain squatting that has yet to be
extensively studied by the research community. We provide the first large-scale,
empirical study to better understand how attackers use combosquatting to perform
a variety of abusive behaviors. Our study examines the lexical characteristics,
temporal behavior, and real world abuse of combosquatting domains. We find that
not only does combosquatting abuse often appear to go unremediated, but its
popularity also appears to be on the rise.

\section{Background}
\label{sec:background}

In this section, we define combosquatting and discuss how it differs from other
types of DNS squatting. Additionally, we discuss how combosquatting is used to
facilitate many different types of abuse. For example, Internet miscreants use
combosquatting to perform social engineering, drive-by-download attacks, malware
communication, and Search Engine Optimization (SEO) monetization. Thus, even
though combosquatting has not been extensively studied, it has far reaching
implications.

\subsection{DNS Squatting \& Combosquatting} 
\label{sub:combosquatting}

Combosquatting refers to the attempt of ``borrowing'' a domain name's reputation
(or brand name) characteristics by integrating a brand domain with other
characters or words. Combosquatting differs from other forms of domain name
squatting, like typosquatting 
and bitsquatting~\cite{nikiforakis2013bitsquatting}, in two fundamental ways:
first, combosquatting does not involve the spelling deviation from the original
trademark and second, it requires the original domain to be {\bf intact} within
a set of other characters. In this paper, we consider a domain name being
combosquatting based on the following definition.

Given the effective second level domain name (e2LD) of a legitimate trademark, a
domain is considered combosquatting if the following two conditions are met: (1)
The domain contains the trademark. (2) The domain cannot result by applying the
five typosquatting models of Wang et al.~\cite{wang2006strider}.

\begin{table}[b]
	\begin{center}
    \begin{tabular}{l l}
      \multicolumn{1}{c}{\textbf{Domain Name}} & \multicolumn{1}{c}{\textbf{Squatting Type}} \\
      \hline 
      youtube[.]com & Original Domain \\
      youtubee[.]com & Typosquatting~\cite{Moore2010} \\
      yewtube[.]com & Homophone-Based Squatting~\cite{nikiforakis2014soundsquatting} \\
      youtubg[.]com & Bitsquatting~\cite{nikiforakis2013bitsquatting} \\
      Y0UTUBE[.]com & Homograph-Based Squatting~\cite{Holgers:2006:CTC:1267359.1267383} \\
      \textbf{youtube-login[.]com} & \textbf{Combosquatting} \\
      \hline 
    \end{tabular}
    \caption{Examples of the different types of domain name squatting for the
      youtube[.]com domain name.}
    \label{tab:squatting-types}
	\end{center}
\end{table}

For example, lets consider the trademark \textit{Example}, such that it is
served by the domain name example[.]com and the e2LD of which is
\textit{example}. Combosquatting domain names, based on this e2LD, could include
any combination of valid characters in the Domain Name System, whether they are
prepended or appended to the e2LD\@. For instance, secure-example[.]com,
myexample[.]com, another-coolexample-here[.]com are cases of combosquatting.
However, wwwexample[.]com and examplee[.]com are not, since they violate the
second clause mentioned earlier. Table~\ref{tab:squatting-types} shows examples
of the different squatting attacks against the youtube[.]com domain name.

\begin{figure*}[t]
  \centering
  \begin{subfigure}[t]{0.30\textwidth}
    \includegraphics[scale=0.15,valign=b]{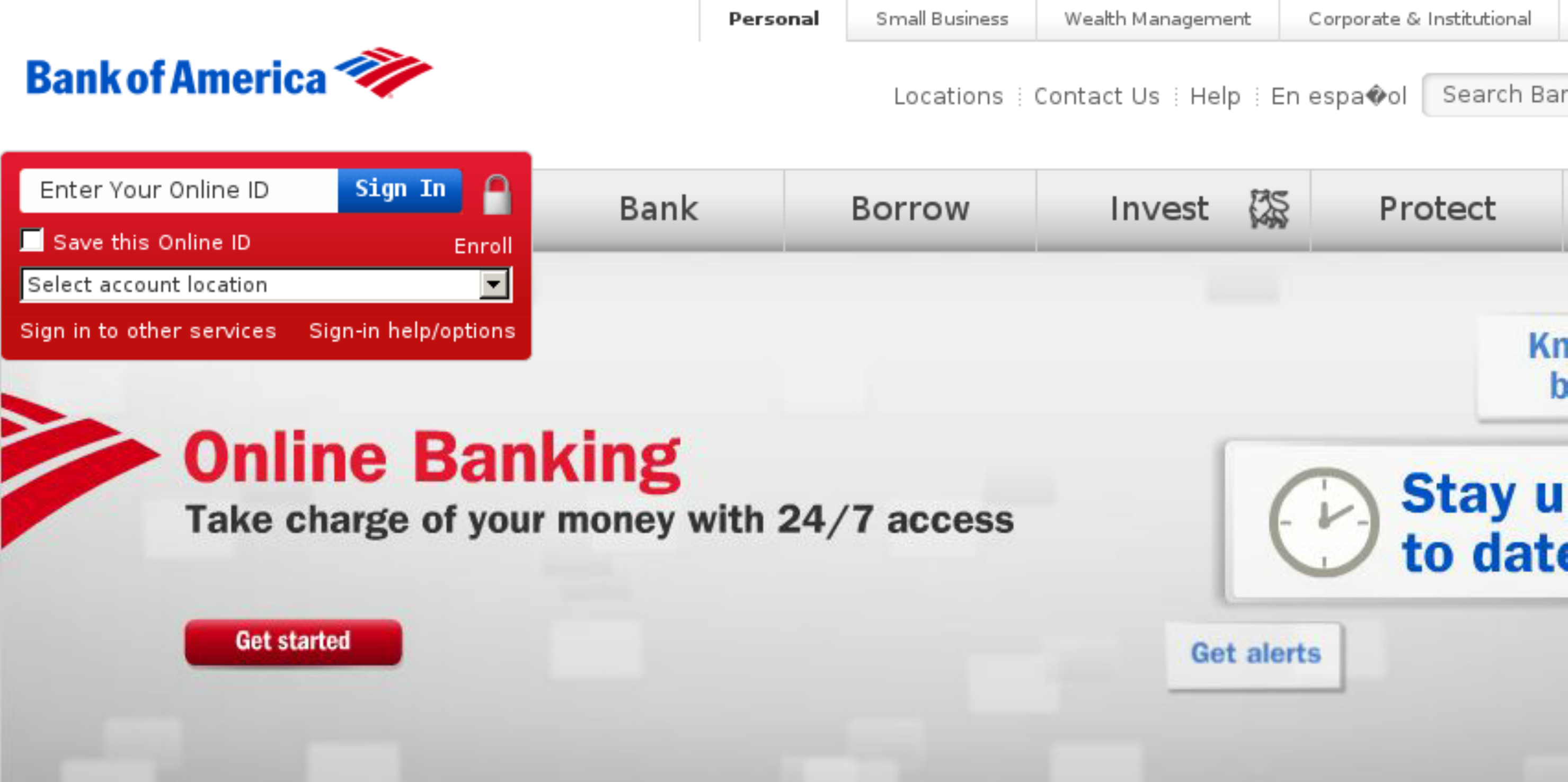}
    \caption{ }
    \label{fig:boa_phish}
  \end{subfigure}
  \begin{subfigure}[t]{0.30\textwidth}
    \includegraphics[scale=0.15,valign=b]{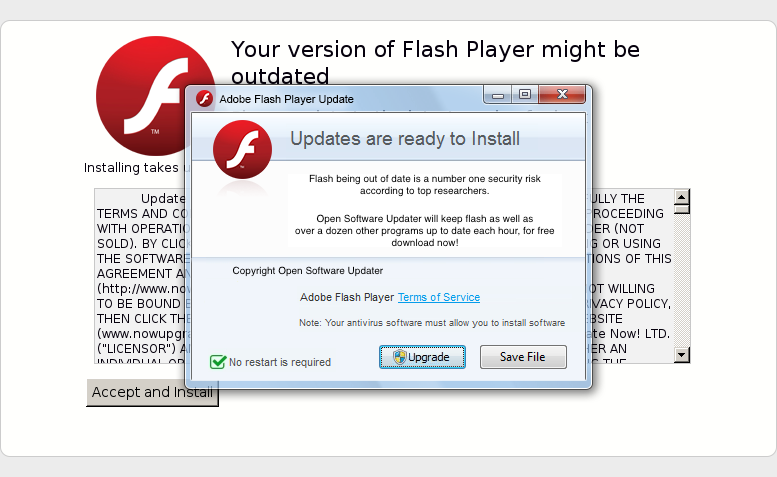}
    \caption{ }
    \label{fig:AirBnB_phish}
  \end{subfigure}
  \begin{subfigure}[t]{0.30\textwidth}
    \includegraphics[scale=0.35,valign=b]{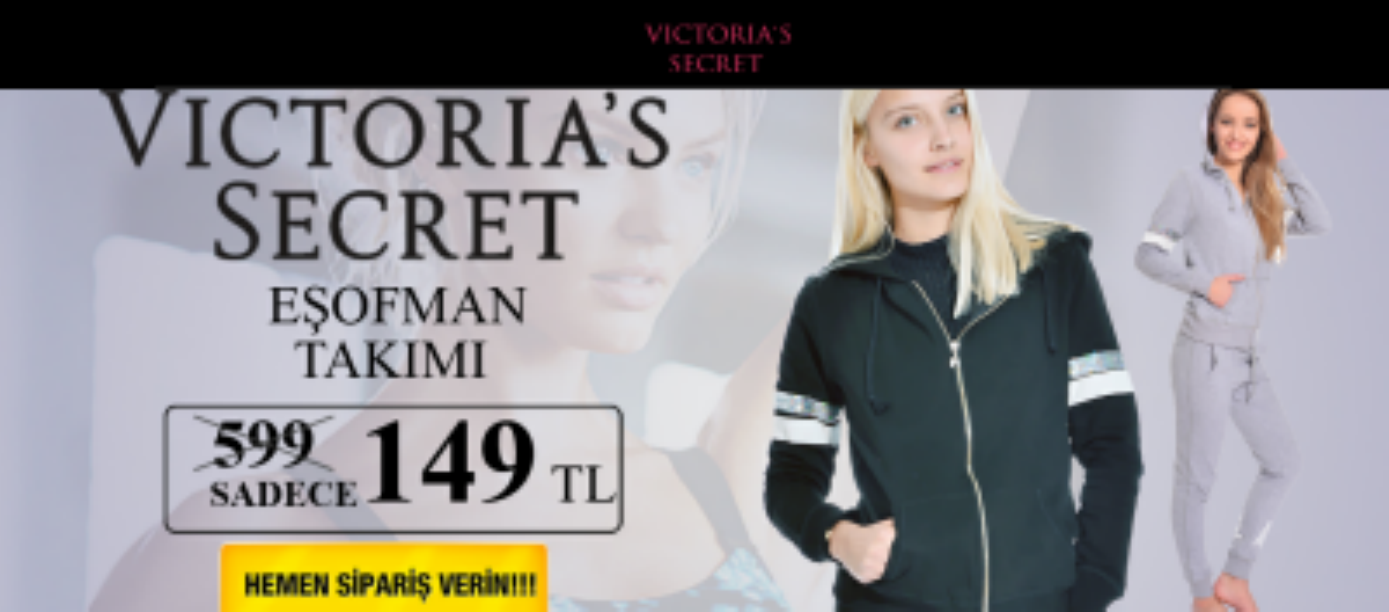}
    \caption{ }
    \label{fig:shady_victoriassecret}
  \end{subfigure}
  \caption{Examples of combosquatting abuse. (a) A typical phishing campaign
    against Bank of America using the domain
    \underline{bankofamerica}-com-login-sys-update-online[.]com. (b) The
    \underline{airbnb}forbeginners[.]com domain uses the AirBnB brand to lure
    users and drop a malware obfuscated as a Flash Update. (c) An example of
    trademark abuse against Victoria's Secret using the domain name
    \underline{victoriassecret}outlet[.]org.}\label{fig:combo_abuse_example}
\end{figure*}

\subsection{Combosquatting Abuse} 

In this section, we discuss the most common types of combosquatting abuse.
Despite common beliefs, combosquatting domains are not only used for trademark
infringement but are also regularly used in a wide variety of abusive
activities---including drive-by downloads, malware command-and-control, SEO, and
phishing. We should note that all cases mentioned next were reported to the
registrars and law enforcement for remediation.

\subsubsection{Phishing}

In generic phishing attacks, where obtaining the user's credentials is the final
goal of the adversary, the attacker would likely register combosquatting domains
close to the targeted organization. For example, in Figure~\ref{fig:boa_phish}
we can see one of those phishing campaigns against Bank of America (BoA) users
that employees the \underline{bankofamerica}-com-login-sys-update-online[.]com
domain. It is worth noting that the phishing page that was hosted on this
combosquatting domain was nearly identical to the actual BoA website. We argue
that this visual similarity, when coupled with the bank's brand name clearly
embedded in the combosquatting domain, makes it highly unlikely that everyday
users of the web would be able to detect this website as phishing.

\subsubsection{Malware}

Delivery of malware and drive-by attacks is another interesting case of
combosquatting abuse. For example, a combosquatting domain can be used to
redirect victims to a page showing fake warnings to lure them into downloading
malicious software. Figure~\ref{fig:AirBnB_phish} shows the domain
\underline{airbnb}forbeginners[.]com being used to lure new AirBnB users. Once
users land on the page, a Flash update request is shown to the end user in what
looks like a Windows dialogue prompt. Thus, the attack attempts to infect the
user by using alerts that suggest Flash Player is outdated and then entice the
user to download a malicious update.

In Table~\ref{tab:malware_domains}, we can see malware related domain names that
were used as Command and Control (C\&C) points for botnets created using popular
malware kits (e.g., Zeus). While it is hard to know for sure why attackers
decide to use domains that contain popular trademarks, such domains could evade
manual analysis of malware communications. The use of combosquatting domain
names is not limited to common malware families, like the ones in
Table~\ref{tab:malware_domains}. As we will see in Section~\ref{sub:data}, using
public reports around targeted attacks and Advance Persistent Threats (APTs), we
identified more than 60 APT C\&C domains that utilize combosquatting, abusing up
to 12 different popular brand names.

\begin{table}[b!]
	\begin{center}
    \begin{tabular}{l l l}
      \multicolumn{1}{c}{\textbf{Domain Name}} & \multicolumn{1}{c}{\textbf{Trademark}}& \multicolumn{1}{c}{\textbf{Abuse Type}} \\
      \hline
      adobejam[.]in & Adobe & Artro C\&C \\
      norton360america[.]biz & Norton & Betabot Botnet\\
      googlesale[.]net & Google & Etumbot \\  
      indexstatyahoo[.]com & Yahoo & Phoenix Kit \\
      pnbcnews[.]ru  & NBC News & Pkybot Botnet \\
      wordpress-cdn[.]org & WordPress & Pkybot Botnet \\
      youtubeee[.]ru & YouTube & Zeus Botnet \\
      google-search[.]ru & Google & Zeus Botnet \\
      \hline
    \end{tabular}
    \caption{Examples of combosquatting domains used by malware as Command and
      Control (C\&C) points.}
    \label{tab:malware_domains}
	\end{center}
\end{table}

\subsubsection{Monetization}

Next to malicious activities mentioned earlier, combosquatting domains have been
heavily exploited in trademark infringement and Search Engine Optimization
(SEO). In this monetization category, the combosquatting domains often advertise
services similar or related to the original services and products offered by the
trademarks being abused. A real world example of such a trademark infringing
domain is presented in Figure~\ref{fig:shady_victoriassecret} in which the
domain name \underline{victoriassecret}outlet[.]org abuses the Victoria's Secret
trademark to offer likely counterfeit products at a lower price.

\section{Measurement Methodology}
\label{sec:methodology}

Measuring the extent of the combosquatting problem is particularly hard because
of the almost unlimited pool of potential domains. However, given the definition
of combosquatting in Section~\ref{sub:combosquatting}, we provide a methodical
way to identify combosquatting domains using various datasets. Additionally, we
discuss our rationale for selecting trademarks that are most likely to be
abused, the type of datasets we use throughout our study, and introduce the
necessary notation utilized from this point on.

\subsection{Trademark Selection}
\label{sub:seed-selection}

While all trademarks could be the subject of combosquatting abuse, it is
arguably not in the best interest of an adversary to use a less known brand for
abuse. In our hypothesis we assume that the adversary would include the
trademark name in the effective second level domain (e2LD) as a way to lure
victims into clicking and interacting with the combosquatting domain and site.

To that extent, we first need to identify the set of popular domains that are
used by major brands (likely to be abused by adversaries). To assemble this list
of domains, we extracted the top 500 domain names in the United States (US) from
Alexa~\cite{alexa}. Our decision to use only the US-centric popular Alexa
domains is due to the underlying datasets we will use for our long-term study
(which are mostly US-centric), as we will see in the following section.

Now, even with the top 500 Alexa list, not all domains are appropriate
candidates for our combosquatting analysis. This is because (1) there are
several brands that employ common words as their brand name and (2) there are
several domains and trademarks that are too short to be considered for
combosquatting. Table~\ref{tab:excluded-trademarks} shows a list of trademarks
that were ignored in the Alexa Top 500 due to the previous considerations.

\begin{table}[t]
 	\begin{center}
    \begin{tabular}{l l l}
      \multicolumn{1}{c}{\textbf{Trademark}} & \multicolumn{1}{c}{\textbf{Domain}} & \multicolumn{1}{c}{\textbf{Potential Squat}} \\
      \hline
      Apple & apple.com & \underline{apple}juice[.]com \\
      AT\&T & att.com & \underline{att}orney[.]com, \underline{att}ack[.]com \\
      Bing & bing.com & plum\underline{bing}[.]com, tu\underline{bing}[.]com \\
      citi (bank) & citi.com & \underline{citi}es[.]com, \underline{citi}zen[.]com \\
      IKEA & ikea.com & b\underline{ikea}ndride[.]com \\
      Cisco & cisco.com & sanfran\underline{cisco}[.]com \\
      \hline 
    \end{tabular}
		\caption{Trademark examples that have been excluded from our study.}
    \label{tab:excluded-trademarks}
  \end{center}
\end{table}

\begin{table*}[t]
  \centering
  \label{tab:original-datasets}
  \begin{tabular}{l r r c l}
    \multicolumn{1}{c}{\textbf{Dataset Type}} & \multicolumn{1}{c}{\textbf{Size}} & \multicolumn{1}{c}{\textbf{Records}} & \multicolumn{1}{c}{\textbf{Time Period}} & \multicolumn{1}{c}{\textbf{Notation}} \\
    \hline
    Passive DNS & 18.1T & $13.1 \times 10^9$ & 2011--01--01 to 2015--10--14 & $PDNS$  \\
    Active DNS & 30.5T & $455 \times 10^9$ & 2015--10--05 to 2016--08--19 & $ADNS$  \\  
    Public BLs & 26.7G & $610 \times 10^6$ & 2012--12--09 to 2016--09--13 & $PBL$  \\  
    APT Reports & N/A & 21,927 & 2008--10--01 to 2016--11--04 & $APT$  \\
    Spamtrap & 35M & 965,911 & 2009--07--17 to 2016--09--13 & $SPA$  \\
    Malware Traces & 34.8G & $1.1 \times 10^9$ & 2011--01--01 to 2016--10--22 & $MAL$  \\  
    Alexa & 42.9G & $1.3 \times 10^9$ & 2012--12--09 to 2016--09--13 & $ALE$  \\
    Certificate Transparency & 842G & $271 \times 10^6$ & 2013--03--25 to 2017--04--13& $CERT$  \\  
    \hline
  \end{tabular}
	\caption{Summary of the raw datasets used in this study.}
  \label{table:DI}
\end{table*}

\begin{table*}[t]
  \centering
  \begin{tabular}{ l r r r r r r r }
    & \multicolumn{3}{c}{\textbf{Passive DNS}} & \multicolumn{3}{c}{\textbf{Active DNS}} & \\
    \multicolumn{1}{c}{\textbf{$\alpha$}} & \multicolumn{1}{c}{\textbf{$\alpha \cap CP$}} & \multicolumn{1}{c}{\textbf{$NoT$}} & \multicolumn{1}{c}{\textbf{$NoC$}} & \multicolumn{1}{c}{\textbf{$\alpha \cap CA$}} & \multicolumn{1}{c}{\textbf{$NoT$}} & \multicolumn{1}{c}{\textbf{$NoV$}} & \multicolumn{1}{c}{\textbf{e2LDs Count}} \\
    \hline
    $CP$ & & & & & & &  2,321,914 \\
    $CA$ & & & & & & & 1,022,083 \\
    \hline
    $C_{mal}$ & 9,283 & 179 & 21 & 6,886 & 174 & 21 & 9,472 \\
    $C_{pbl}$ & 3,750 & 135 & 21 & 4,787 & 128 & 21 & 5,844 \\
    $C_{apt}$ & 59 & 11 & 8 & 56 & 12 & 8 & 65 \\
    $C_{spa}$ & 2,296 & 126 & 20 & 6,400 & 148 & 20 & 6,400 \\
    $C_{abuse}$ & 14,965 & 201 & 21 & 17,586 & 200 & 21 &  21,173 \\
    \hline
    $C_{ale}$ & 45,619 & 244 & 22 & 37,098 & 244 & 22 &  48,197 \\
    \hline
  \end{tabular}
  \caption{The combosquatting datasets, and their relational statistical
    properties. $NoT$: Number of unique trademarks in a set of domains and
    $NoC$: Number of unique business categories in a set of domains.
    $C_{abuse}=\{ C_{mal} \cup C_{pbl} \cup C_{apt} \cup C_{spa}\}$.}
  \label{table:DS}
\end{table*}

We manually inspected all 500 top Alexa domains to exclude domains that fall
into the two aforementioned categories. The remaining set contains 246 domains
that we will consider in our combosquatting study. We will refer to this list of
domains as \textit{seed} throughout the rest of the paper. The trademarks
selected belong to companies that are active in different business categories.
Thus, we are able to group them together into 22 categories based on the type of
services/products they offer.

We derived this categorization using the Alexa list~\cite{alexa}, the
TrendMicro~\cite{trendmicrocheck} website and the DMOZ database~\cite{dmoz}. We
manually verified the categories and merged any differences between the
platforms to create a consistent list. The vast majority of the domains had a
stable Alexa rank over time. At the same time, we added seven domains that were
a priori chosen in the ``Politics'' category and 15 for the ``Energy'' category,
following the same process as before. We manually included the energy sector
because it is part of the critical infrastructure and the politics because of
the US Presidential elections of 2016.

\subsection{Datasets}
\label{sub:data}

Since our goal is to study combosquatting both in depth and over time, we
require a variety of different datasets. Table~\ref{table:DI} summarizes the raw
datasets used in this study, and Table~\ref{table:DS} lists the most important
relationships between them. We provide more detail about each of these
datasets below. \\

\noindent{\bf Passive DNS:} The passive DNS dataset ($PDNS$) consists of DNS
traffic collected since 2011, above a recursive DNS server located in the
largest Internet Service Provider (ISP) in the US. Specifically, this dataset
contains the DNS resource records (RRs) from all successful DNS resolutions
observed at the ISP, including their daily lookup volume. \\

\noindent{\bf Active DNS:} We also utilize an active DNS ($ADNS$) dataset, which
we obtain daily from the Active DNS project~\cite{activedns}. Since the duration
of this dataset is less than a year, it does not have a complete temporal
overlap with our $PDNS$ dataset. While we will use the $PDNS$ and $ADNS$
datasets for most measurement tasks, we will also use a variety of smaller
datasets to label and measure abuse in these combosquatting datasets. Again, in
Table~\ref{table:DI} we can see these five different datasets used in
this study. \\

\noindent{\bf Public Blacklists:} We collect historic public blacklisting
($PBL$) information about domains that have been identified by the security
community as abusive and placed in various public lists~\cite{pbl_abusech,
  pbl_malwaredomainlist, pbl_malc0de, pbl_sagadc, pbl_hphosts, pbl_sans,
  pbl_itmate, pbl_driveby}. These blacklists have been collected from 2012 until
2016 and overlap with our passive and active DNS datasets. \\

\noindent{\bf Advanced Persistent Threats:} Using public Advanced Persistent
Threat ($APT$) reports~\footnote{\url{http://tinyurl.com/apt-reports}}, we
manually extract and verify domain names used in such documented attacks (APT). \\

\noindent{\bf Spam Trap:} A security company provides us with spam
trap~\cite{kreibich2008spam} data that is labeled using their proprietary
detection engine ($SPA$). \\

\noindent{\bf Malware Feeds:} The same security company and a university
provides us with two feeds of domains from dynamic execution of malware samples
since 2011 ($MAL$). \\

\noindent{\bf Alexa List:} To eliminate potentially wrong classification of a
domain as abusive (false positive) in the aforementioned datasets, we create a
``whitelist'' based on the Alexa list. We take the domains that appeared in the
top 10,000 of the Alexa list for more than 90 consecutive days in the last five
years and create a set of domains as indicators of benign activity ($ALE$). \\

\noindent{\bf Certificate Transparency:} Google's Certificate Transparency
(CT)~\cite{certificate-transparency} project provides publicly auditable,
append-only logs of certificates with cryptographic properties that can be used
to verify the legitimacy of certificates seen in the wild. The official CT
website provides a list of known, active logs that can be publicly crawled. We
used this list to download all records from those logs up to April 13, 2017.
This resulted in a dataset of approximately 271M certificates.

\subsection{Linking Datasets}
\label{sub:linking-datasets}

Next, we project the selected trademarks, into the raw datasets presented in
Table~\ref{table:DI}, and derive the trademark--specific datasets, which can be
seen in Table~\ref{table:DS}. The datasets in Table~\ref{table:DS} will be used
to study the combosquatting problem in depth since 2011. We begin by extracting
the Combosquatting Passive ($CP$) and Combosquatting Active DNS ($CA$) set of
domains, which reflect combosquatting domains containing at least one of the
trademarks of interest in the Passive and Active DNS datasets, respectively. The
cardinalities of these two sets are of the order of millions of domain names
(2.3M for the $CP$ set and 1M for the $CA$), and all combosquatting domain abuse
should be bounded by the size of the two sets.

Following the same process, we identify the combosquatting domains in the PBL,
APT, Spamtrap, Malware and Alexa sets, deriving $C_{pbl}$, $C_{apt}$, $C_{spa}$,
$C_{mal}$ and $C_{ale}$, respectively. The cardinalities of these sets can be
seen in Table~\ref{table:DS} where they span from a few domains ($C_{apt}$) to
several tens of thousands of domains ($C_{mal}$ and $C_{ale}$). Finally, we will
define $C_{abuse}$ as the set of domains in all malicious categories of
combosquatting domains, namely $C_{pbl}$, $C_{apt}$, $C_{spa}$, and $C_{mal}$.

\section{Measuring Combosquatting Domains}
\label{sec:measurements}

In this section we present short and long term measurements revolving around the
combosquatting domains in our datasets. We begin by investigating the
differences between typosquatting and combosquatting. At the same time we
discuss which words attackers choose to combine with popular trademarks more
frequently. Then, we study the temporal properties of the domain names in the
combosquatting passive and active DNS datasets. This analysis will help us
understand how these combosquatting domains evolved since 2011.

In particular, we observe that the number of combosquatting domain names in our
passive and active DNS datasets are steadily increasing; in contrast, the
domains in the $C_{abuse}$ set remain stable over time. At the same time, we
observer that the security community is lagging behind the detection of
malicious combosquatting domains, in many cases up to several months, despite
being an obvious target of abuse. Finally, we provide an analysis of the DNS and
IP hosting infrastructure that combosquatters tend to employ. The domains in the
$C_{abuse}$ set tend to utilize significantly more agile hosting infrastructure,
which could be used as a signal to identify abusive combosquatting domains on
the rise.

\begin{figure*}[t]
  \begin{center}
    \includegraphics[scale=0.5]{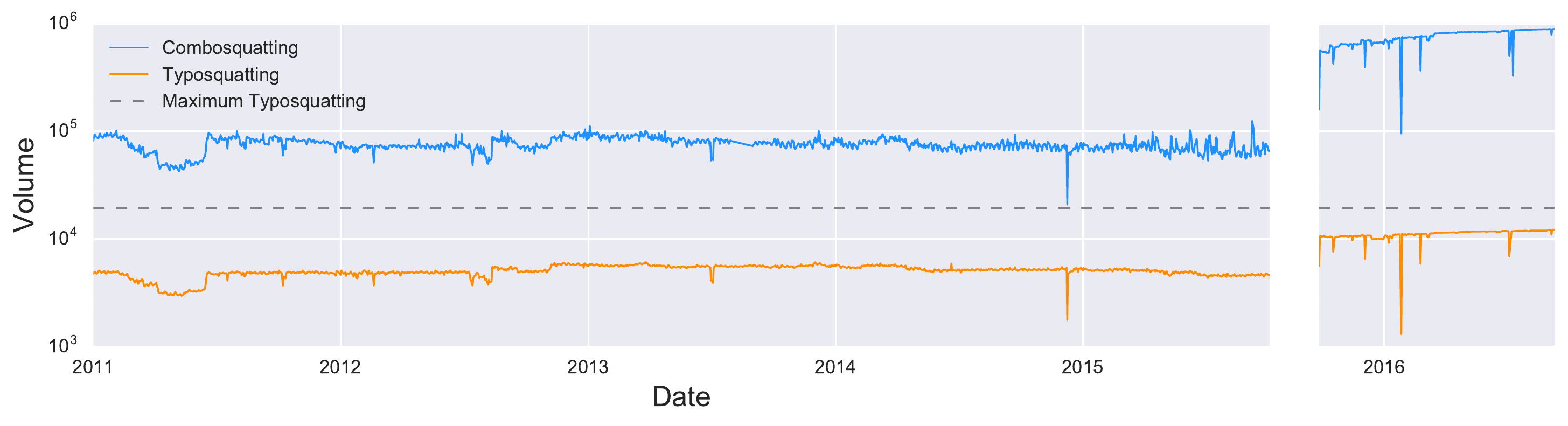}
    \caption{Number of active Combosquatting and Typosquatting domain names per
      day. The left hand side part of the plot depicts the passive DNS period,
      whereas the right one reflects domains found in the active DNS dataset.}
    \label{fig:domain_number}
  \end{center}
\end{figure*}

\subsection{Combosquatting versus Typosquatting}
\label{sec:combo-vs-typo}

Since typosquatting is, by far, the most researched type of domain squatting, we
begin our discussion of combosquatting by comparing it with typosquatting.
Figure~\ref{fig:domain_number} shows the number of active typosquatting and
combosquatting domains targeting our evaluated trademarks since 2011. To
identify typosquatting domains, we use the five typosquatting models of Wang et
al.~\cite{wang2006strider} to generate all possible typosquatting domains and
search for those domains in our DNS datasets. The left part of the plot is based
on our passive DNS dataset while the right part is based on the active DNS
dataset. One can clearly see that, even though combosquatting has evaded the
attention of researchers, it is significantly more prevalent than typosquatting,
with the number of daily combosquatting domains being almost two orders of
magnitude larger than the number of typosquatting domains.

In comparison with other types of domain squatting phenomena such as
typosquatting, combosquatting has a unique property in that it lacks a
generative model. For all other types of domain squatting, researchers can start
with an authoritative domain, and by performing character and bit swaps, they
can exhaustively list the possible squatting permutations for a given type of
domain squatting. For example, the dotted line in Figure~\ref{fig:domain_number}
indicates the maximum number of typosquatting domains possible when considering
the evaluated trademarks and typosquatting models~\cite{wang2006strider}. In
combosquatting, however, attackers are free to prefix and postfix a trademark
with one or more keywords of their choice, bounded only by the maximum number of
characters allowed for any given label by the DNS
protocol~\cite{rfc1034,rfc1035}.

Another difference that is closely related to the lack of a generative model, in
terms of attack scenarios, has to do with the way attacks are rendered.
Typosquatting can be a passive attack for the adversary, who simply must wait
until a user accidentally types in a domain. However, combosquatting requires
more active involvement from the attacker because, while a user may accidentally
type paypa[.]com instead of paypal[.]com, an attacker cannot register
paypal-members[.]com and reasonably expect users will accidentally type those
eight extra characters. Therefore, miscreants that rely on combosquatting must
coerce users (e.g. via spam emails and social networks) to visit combosquatting
domains.

To increase the chances that users will interact with their malicious
combosquatting domains, attackers can use services like \textit{Let's
  Encrypt}~\cite{letsencrypt} to both freely and automatically obtain TLS
certificates for their domains. In fact, Let's Encrypt has recently come under
criticism for choosing to eschew any sort of security checks before giving
domain owners a TLS certificate~\cite{cas-role-malware}. To quantify the
frequency with which attackers obtain certificates for their malicious domains,
we searched the 271 million certificates obtained via the Certificate
Transparency append-only log (described in Section~\ref{sub:data}) and
discovered that 691,182 certificates were given to a total of 107,572
fully-qualified combosquatting domains related to our trademarks, since 2013,
with 41.5\% of the certificates being issued by Let's Encrypt. In contrast, only
3,011 certificates were issued for typosquatting domains. This finding further
confirms the intuition that typosquatting and combosquatting are two distinct
phenomena with different threat models and attack strategies.

In summary, we argue that existing domain squatting detection systems are not
taking combosquatting domains into account (since they cannot generate them) and
combosquatting requires its own analysis due to the scale of the problem and the
different threat models involved.

\subsection{Lexical Characteristics}
\label{sec:lexical}

The lack of generative models for combosquatting, makes it hard to proactively
create and evaluate domains. Therefore, we utilize the DNS datasets mentioned
previously, to identify combosquatting domains and analyze their composition. In
particular, we see that adversaries do not usually register lengthy domains and
do not use many words when generating the domains. We also find that there are
certain words that adversaries favor when generating abusive combosquatting
domains. Some words are independent of the trademark's business category, and
other words are specific to a single category.

\begin{figure*}[t]
  \centering
  \begin{subfigure}[t]{0.32\textwidth}
    \includegraphics[scale=0.32,valign=b]{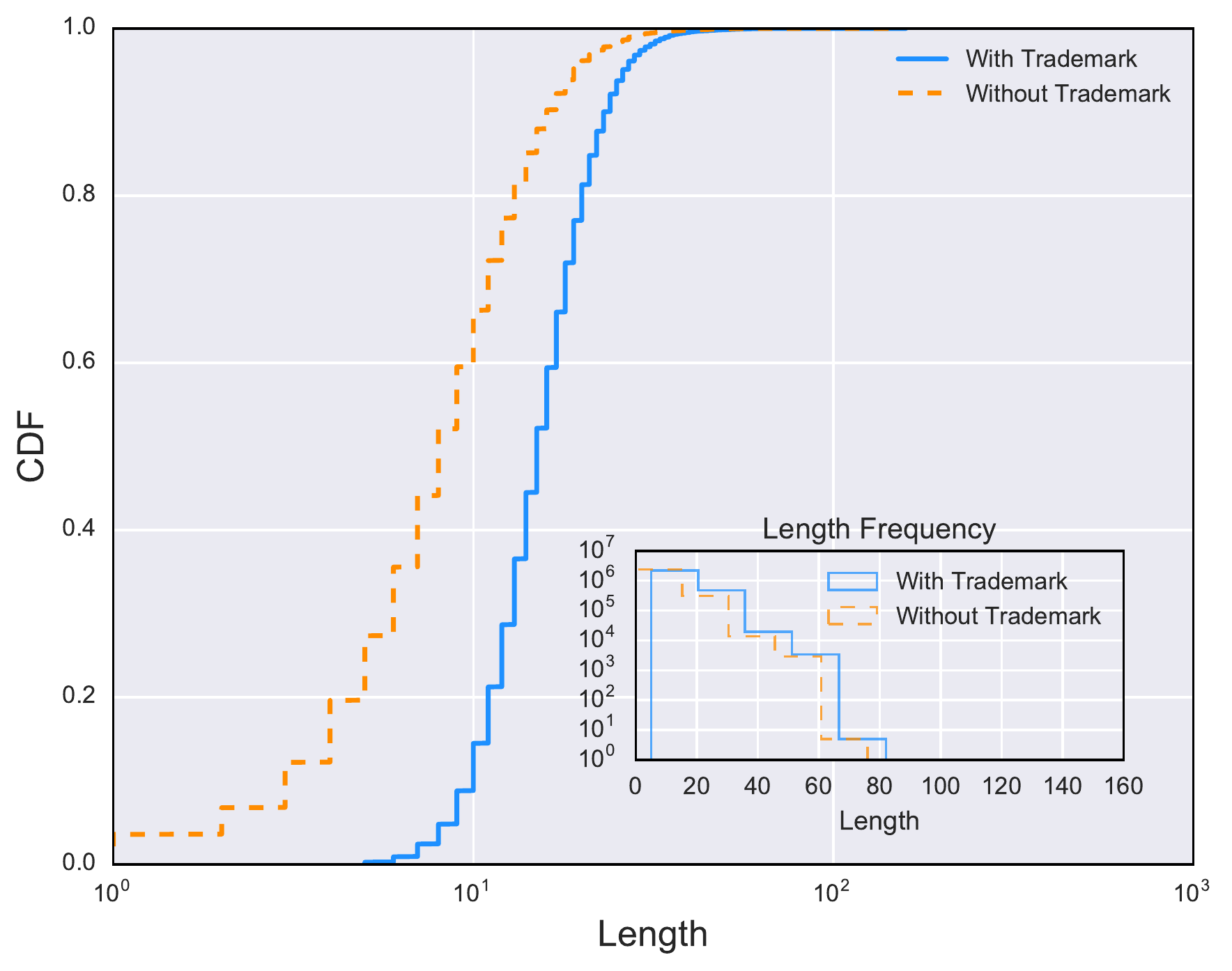}
    \caption{ }
    \label{fig:combosquatting-length}
  \end{subfigure}
  \begin{subfigure}[t]{0.32\textwidth}
    \includegraphics[scale=0.32,valign=b]{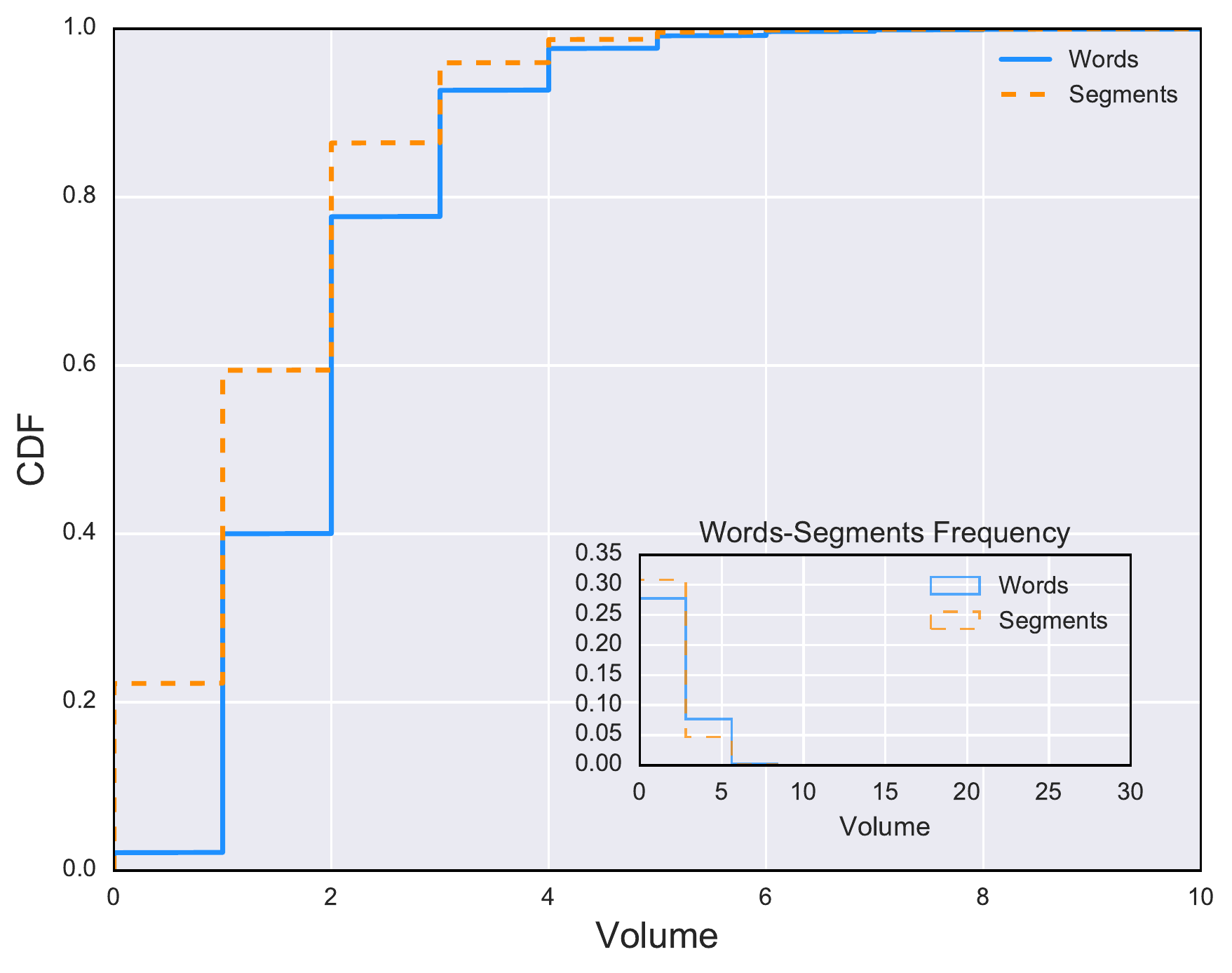}
    \caption{ }
    \label{fig:words_per_trademark_cdf}
  \end{subfigure}
  \begin{subfigure}[t]{0.32\textwidth}
    \includegraphics[scale=0.18,valign=b]{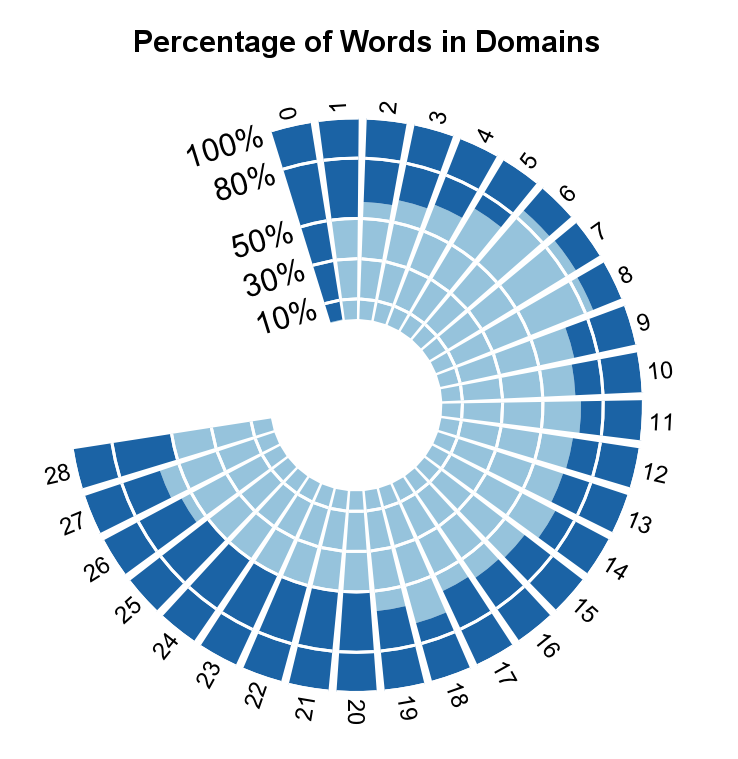}
    \caption{ }
    \label{fig:words-radial}
  \end{subfigure}
  \caption{Lexical Characteristics of combosquatting domains. (a) Length of the
    Combosquatting domain names, including and excluding the original trademark.
    (b) CDF of the number of segments and words. We limit the x-axis of the
    outer plot for the sake of readability. (c) Number of segments used in
    combosquatting domain names. For each number of segments the percentage of
    English words is presented in blue color.}
  \label{fig:lexical_char}
\end{figure*}

\begin{figure*}[t]
  \begin{center}
    \includegraphics[scale=0.50]{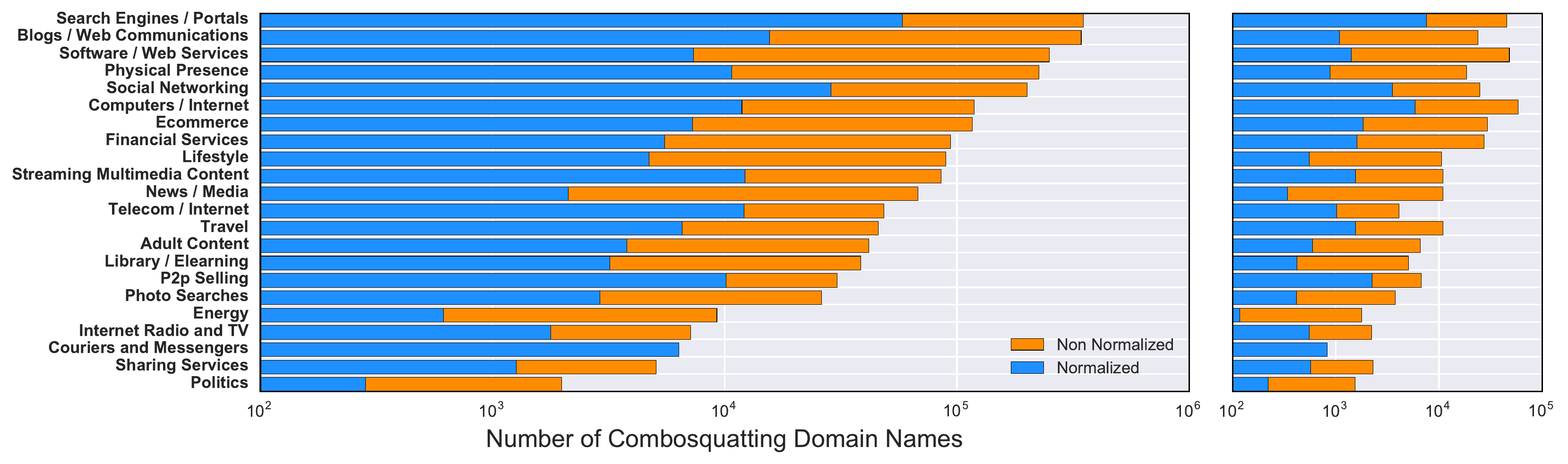}
    \caption{Normalized and absolute size of the combosquatting domains in our
      datasets per business category.}
    \label{fig:squatting-per-category}
  \end{center}
\end{figure*}

Figure~\ref{fig:combosquatting-length} shows the Cumulative Distribution
Function (CDF) of the length of all identified combosquatting domains. There we
can see that even though an attacker can, in principle, construct very long
domains, 60\% of the identified combosquatting domains were using less than ten
characters and 80\% of the combosquatting domains were using less than 22
characters (excluding the original squatted trademark). This provides an early
indication that the vast majority of the attackers carefully construct
combosquatting domain names without attempting to reach the limits afforded to
them by the DNS protocol.

To better understand the construction of combosquatting domains, we extract the
non-Top Level Domain (non-TLD) part of each domain (e.g. we extract
\texttt{facebookfriends} from \texttt{facebookfriends [.] com}) and use the
\textit{word segmentation} algorithm described in~\cite{segaran2009beautiful}.
This algorithm takes a string as input and outputs sequences of that string that
have a high probability of being standalone tokens, along with a confidence
score for the provided tokenization.

We validate the output tokens provided for each combosquatting domain against
four dictionaries: (1) the PyEnchant \texttt{en\_US} Python
dictionary~\cite{pyenchant} to identify English words, (2) the \textit{No
  Swearing} dictionary~\cite{noswearing} to identify swearing-related words, and
both (3) the SWOPODS~\cite{sowpods} and (4) \textit{No Slang}~\cite{noslang}
dictionaries to identify slang words in US English. Tokens that are found in any
of these dictionaries are referred to as \textit{words} and, when not found, we
simply call them \textit{segments}.

Figure~\ref{fig:words_per_trademark_cdf} depicts a CDF of the number of tokens
and number of words that were identified for each domain. We see that almost
80\% of the domains have at most two dictionary-words present, and 90\% have at
most three words. At the same time, we have found a limited number of cases that
contain up to 28 words and segments. These results validate our earlier
length-based claim that squatters appear to be methodical in their construction
of combosquatting domain names. We note that stop words and other short words
have not been removed from our datasets because they are frequently used by
combosquatting domains.

Figure~\ref{fig:words-radial} shows the correlation of segments (cyan) and
actual words (blue). Every bin in the radial histogram represents the number of
tokens identified in each domain. The presented percentage captures the number
of actual words versus segments that we were able to distinguish. As we can see,
the middle ranges of token counts (6 to 19) have a lot more segments than words,
whereas when the domain consists of fewer tokens, the number of words found in
the dictionaries mentioned earlier increases. On average, half of the tokens are
words and the other half are segments. This is likely an artifact of the
attackers' attempts to register domains that might include typos or several
strings close to words, which could be overlooked by the targets, in order to
increase their arsenal of combosquatting domains. Consider, for example, the
following list of domain names that we identified as combosquatting and all
point to a credit card activation campaign.

\begin{verbatim}
activatemycrbankofamerica[.]com
activatemycrebankofamerica[.]com
activatemycredbankofamerica[.]com
activatemycredibankofamerica[.]com
activatemycreditbankofamerica[.]com
activatemycreditcabankofamerica[.]com
activatemycreditcarbankofamerica[.]com
activatemycreditcardbankofamerica[.]com
\end{verbatim}

In terms of the words that attackers combine with abused trademarks, the top
twenty words across all trademark categories were: \textit{free, online, code,
  store, sale, air, best, price, shop, head, home, shoes, work, www, cheap, com,
  new, buy, max,} and \textit{card}. Since the top twenty words represent all of
our 22 categories, they include terms that can be found either in one or
multiple trademark categories. For example, the word ``free'' can be found in 12
of the 22 categories, suggesting that attackers commonly combine the word
``free'' with popular trademarks associated with paid goods (such as shopping,
movies, and TV shows) to lure users into interacting with their websites.
Contrastingly, certain words appear in a single category of trademarks, such as
``cheap'' which is found only in the online shopping category.

Due to space limitations, we make Table~\ref{table:words} available in the
Appendix that presents the ten most frequent words for each trademark category.
We see that many of the popular words closely correlate with the type of
trademark being abused, like the words \textit{apple, game} and \textit{phones}
being popular in the ``Computers/Internet'' category and the words
\textit{president, vote,} and \textit{elect} being popular in the ``Politics''
category. The word selection by the adversaries clearly indicates that most
registered combosquatting domains have been carefully constructed to match the
expected context of each abused trademark. This is a property unique to
combosquatting, since any other type of squatting is bounded to the squatted
domain name itself. For example, the search space in typosquatting, from which
adversaries can choose domain names is bounded to the length of the domain and
the characters used, limiting the agility and multiformity of the threat.

\subsection{Temporal Analysis}
\label{sec:temporal}

\begin{figure*}[t]
  \centering
  \begin{subfigure}[t]{0.32\textwidth}
    \includegraphics[scale=0.32,valign=b]{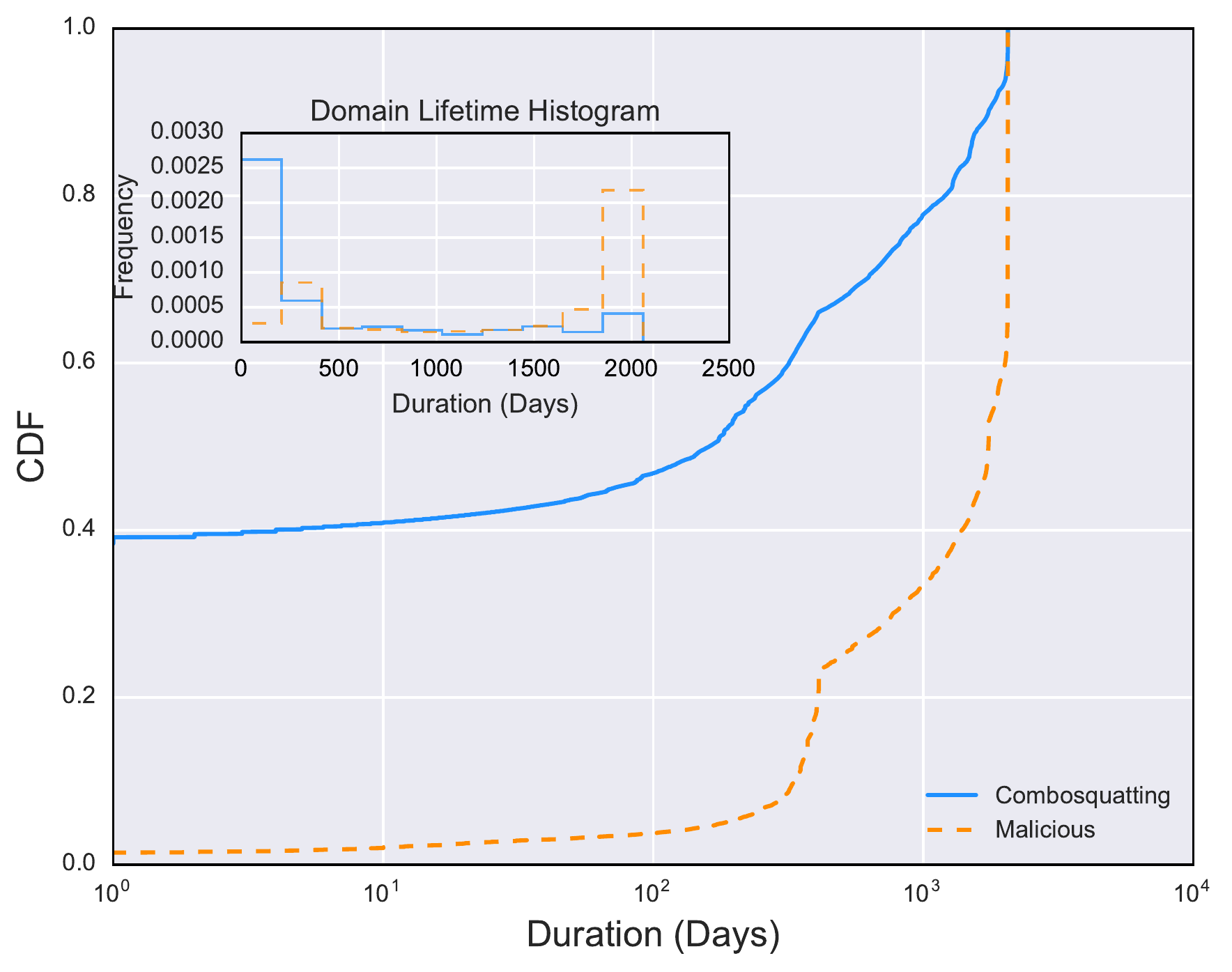}
    \caption{ }
    \label{fig:lifetime}
  \end{subfigure}
  \begin{subfigure}[t]{0.32\textwidth}
    \includegraphics[scale=0.32,valign=b]{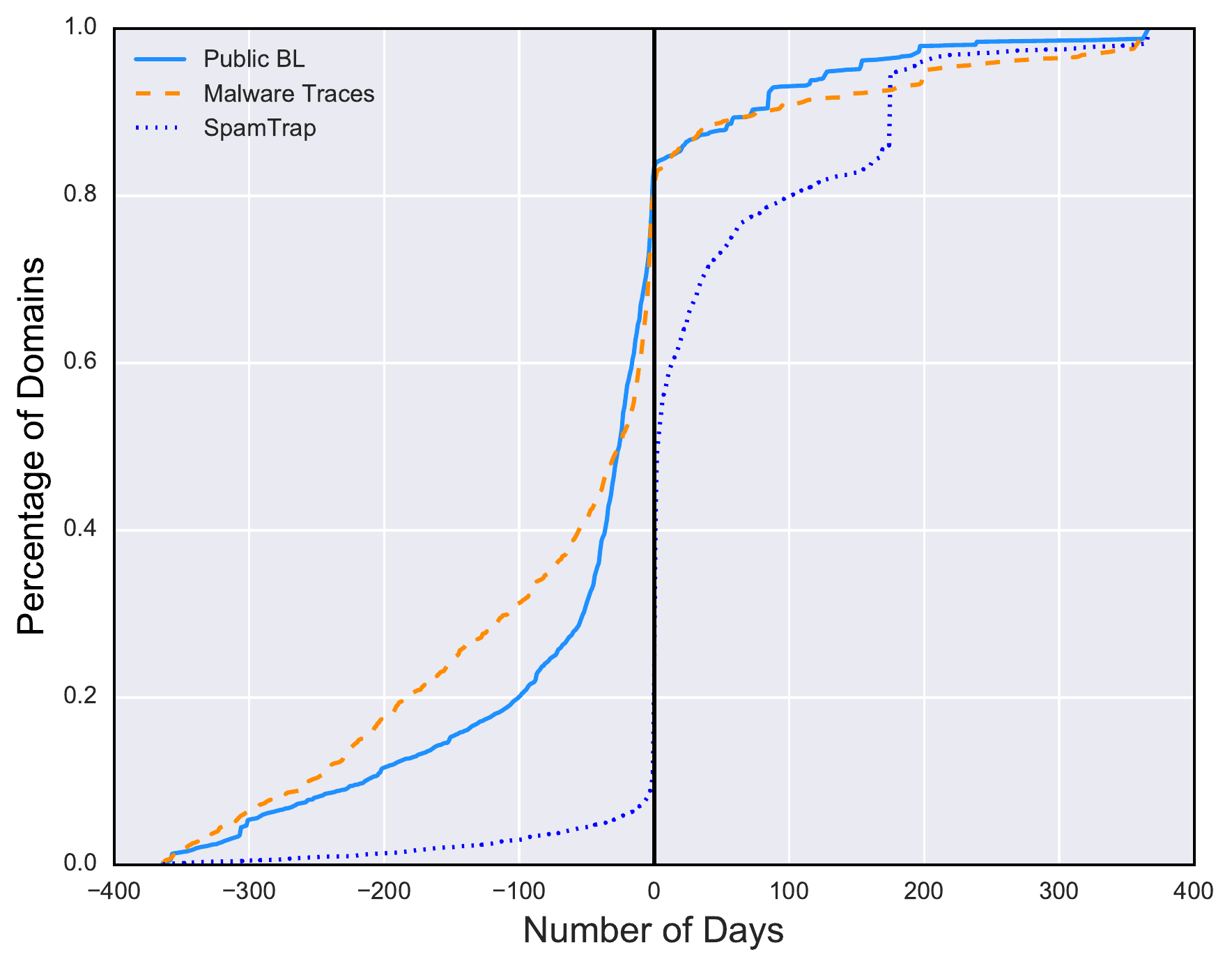}
    \caption{ }
    \label{fig:pbl_malware_time_diff}
  \end{subfigure}
  \begin{subfigure}[t]{0.32\textwidth}
    \includegraphics[scale=0.32,valign=b]{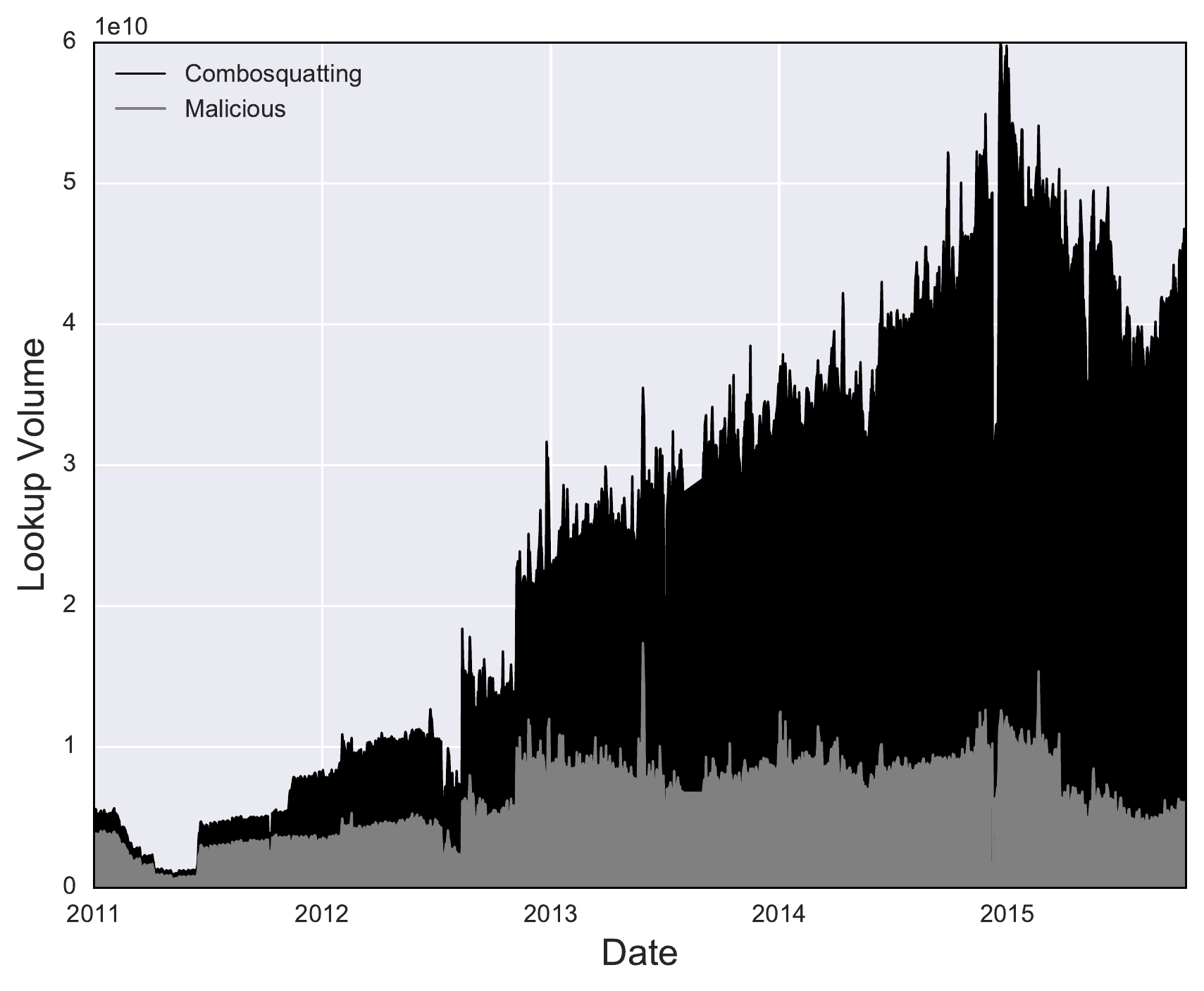}
    \caption{ }
    \label{fig:overtime_lookups}
  \end{subfigure}
  \caption{Infrastructure characteristics of combosquatting domains. (a) A CDF
    of the domain name lifetime in the $CP$ set. (b) The difference between the
    time a combosquatting domain name was first seen in our datasets and the day
    it first appeared in a Public Blacklist, the Malware Traces dataset, or the
    security vendor's spam trap. The plot shows the cumulative volume of domains
    over time, normalized by the maximum number of domains in each dataset. (c)
    The DNS lookup volume for the domain names in the CP set vs. the malicious
    ($C_{abuse}$) domains.}\label{fig:infra_char}
\end{figure*}

\begin{figure}[t]
  \begin{center}
    \includegraphics[scale=0.32]{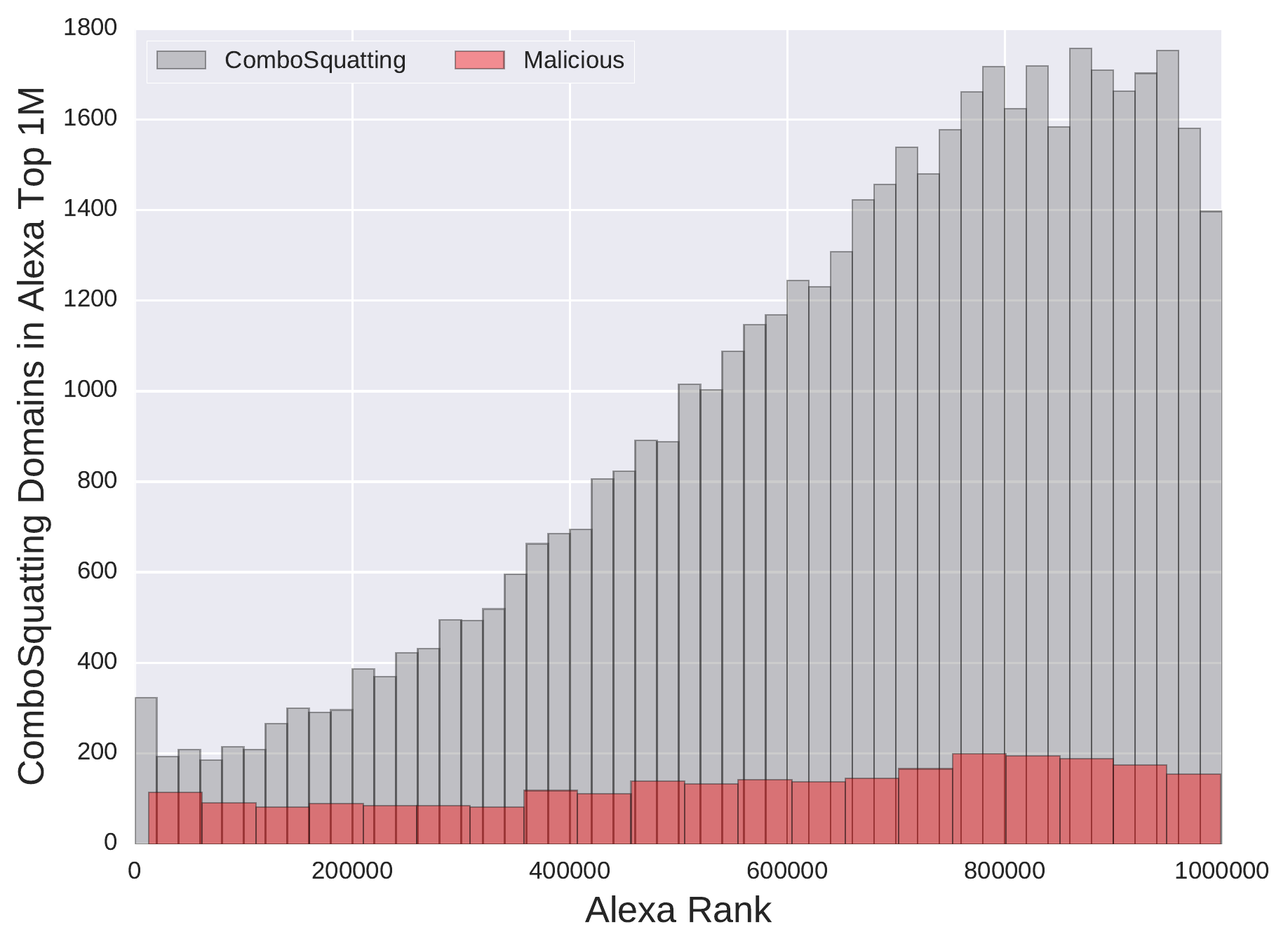}
    \caption{Distribution of the Alexa ranks for combosquatting domains since
      2011. The plot depicts the mean rank for the domain names over the period
      of our $C_{ale}$ dataset.}
    \label{fig:alexa_histogram}
  \end{center}
\end{figure}

\begin{figure*}[t!]
  \centering
  \begin{subfigure}[t]{0.32\textwidth}
    \includegraphics[scale=0.32,valign=b]{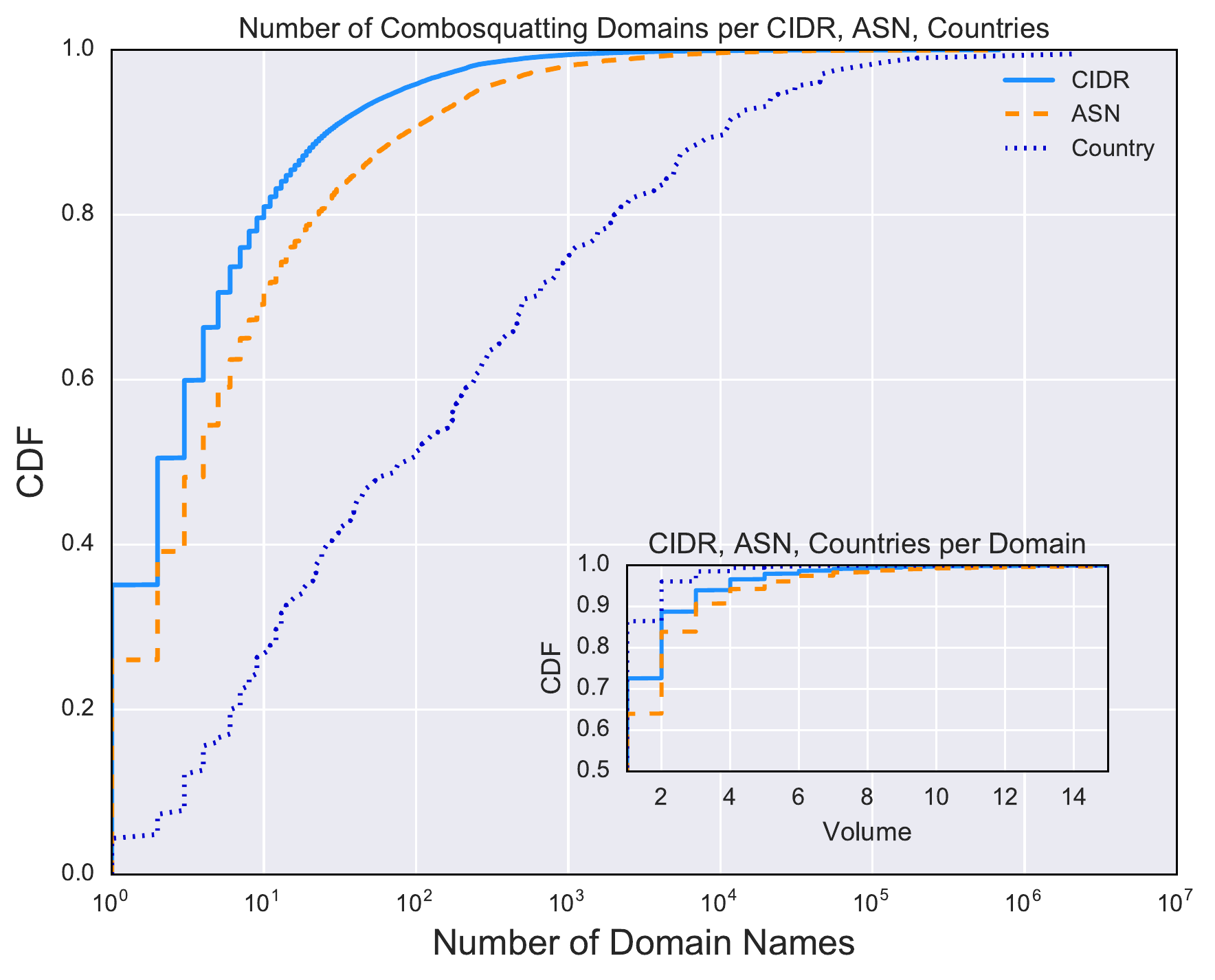}
    \caption{ }
    \label{fig:cp_ac_infra}
  \end{subfigure}
  \begin{subfigure}[t]{0.32\textwidth}
    \includegraphics[scale=0.32,valign=b]{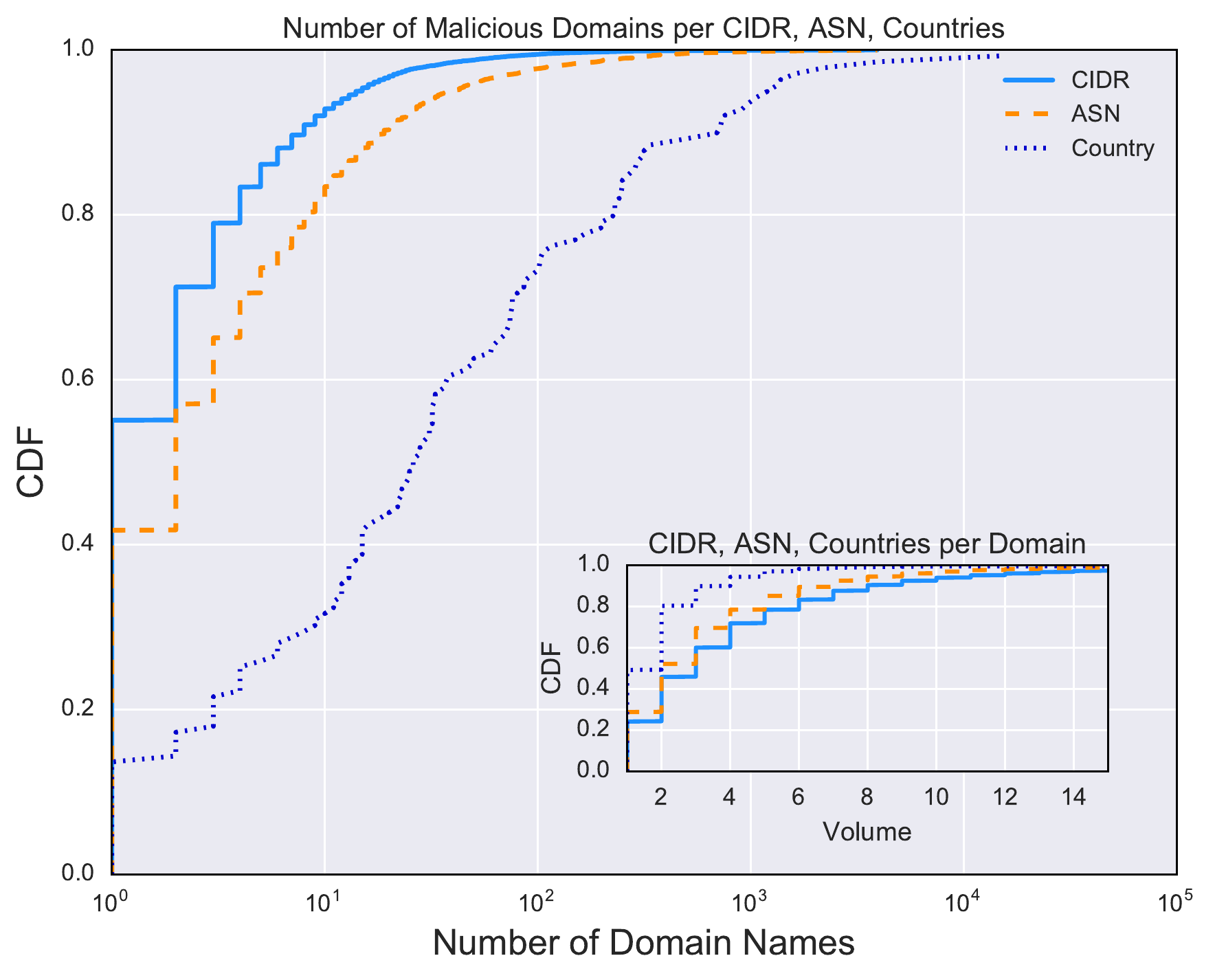}
    \caption{ }
    \label{fig:mal_infra}
  \end{subfigure}
  \begin{subfigure}[t]{0.32\textwidth}
    \includegraphics[scale=0.32,valign=b]{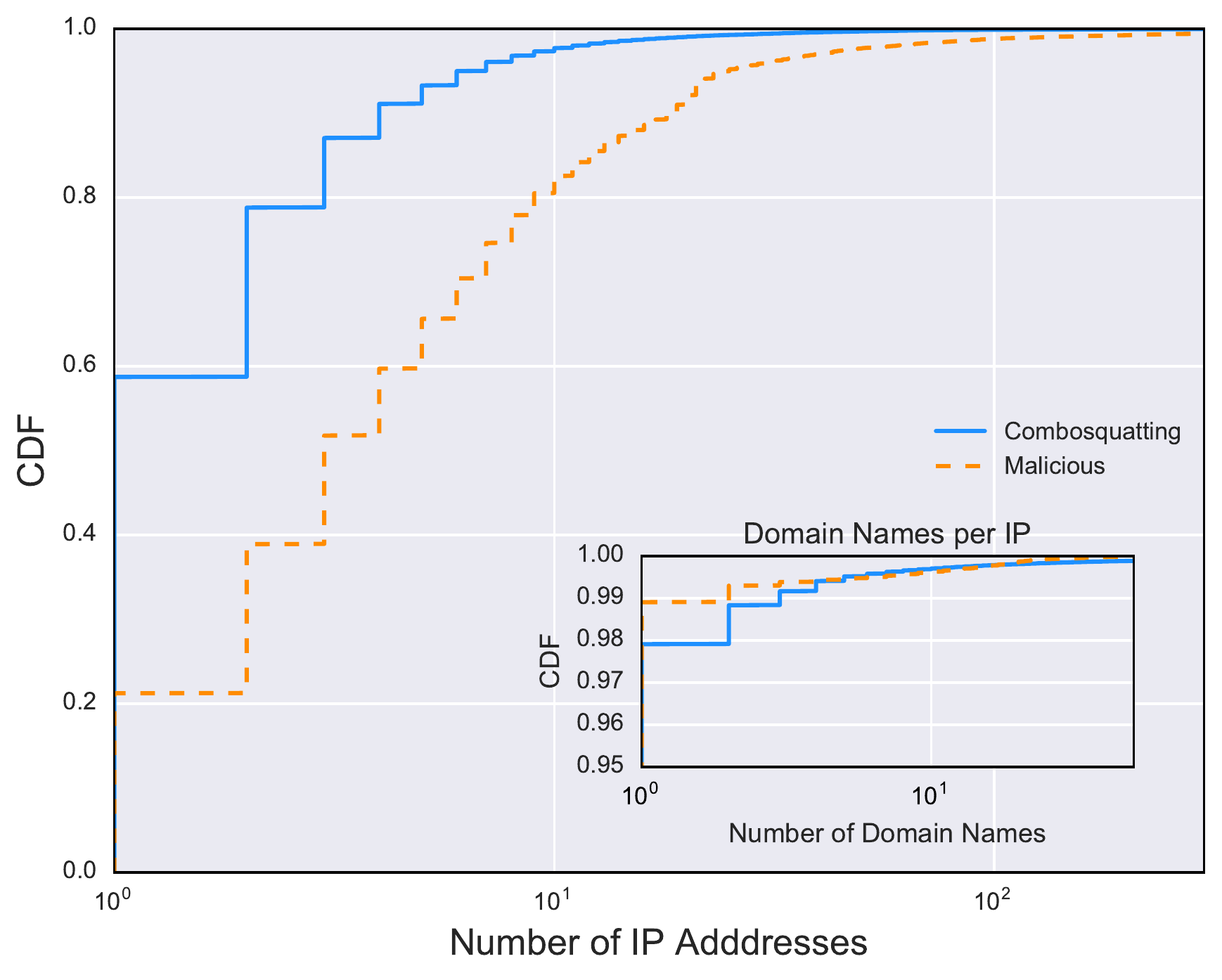}
    \caption{ }
    \label{fig:ips}
  \end{subfigure}
  \caption{Infrastructure distributions for combosquatting Domains. (a) Number
    of combosquatting domains per CIDR, ASN, and Country for all combosquatting
    domain names. The inset plot shows the CIDR, ASN, Country Code frequency
    distribution per combosquatting domain in the $CP$ and $CA$ sets. (b) Number
    of malicious domains ($C_{abuse}$) per CIDR, ASN, Countries. The inner plot
    shows CIDR, ASN, Countries per malicious ($C_{abuse}$) combosquatting
    domain. (c) CDFs for the number of IP addresses that domains in the
    combosquatting ($CP$ and $CA$) and malicious ($C_{abuse}$) utilize during
    their lifetime.}
\end{figure*}

In Section~\ref{sub:seed-selection} we presented our process for selecting the
trademarks we use in our study, and in Section~\ref{sub:data} we discussed the
different datasets we use to measure the phenomenon. Using these trademarks and
the dataset notation from Table~\ref{table:DS}, we study the temporal properties
of combosquatting domains since 2011. We find that clients are increasingly
resolving combosquatting domains and that more than half of all combosquatting
domains share a minimum lifespan of at least three months; in contrast, the
majority of abusive domains are active for more than a year. We also see that
malicious domains appear in the DNS datasets several months before they appear
in our abusive dataset and they even make it into the top thousands ranks in the
Alexa list.

Figure~\ref{fig:squatting-per-category} shows the number of combosquatting
domain names we were able to place in the passive (left) and active (right) DNS
datasets. The orange color represents the total number of combosquatting domain
names we are able to identify in our datasets for each of the trademark
categories. Blue shows the normalized number based on the number of trademarks
that appeared in each category. While most of the combosquatting domain names
are in ``\texttt{Information Technology}'' related categories, our dataset is
not biased, as the sets $CP$ and $CA$ contain a significant number of domains
across all trademarks and business categories.

By focusing our attention on the combosquatting passive DNS set, we can see the
days in which a combosquatting domain name is available in our datasets.
Figure~\ref{fig:lifetime} shows the CDF of this lifetime of the domains in the
$CP$ set. We measure the lifetime of a combosquatting domain as the number of
days between the first and last time we saw it appearing in our passive DNS
dataset. Almost 50\% of the domain names in the CP set were active for at least
100 days. In the same figure, we can observe the malicious class of
combosquatting domain names, which are in the $C_{abuse}$ set (presented earlier
in Table~\ref{table:DS}).

Interestingly, Figure~\ref{fig:lifetime} also shows that the lifetime of abusive
combosquatting domains is greater than the entire combosquatting passive DNS
set. This makes intuitive sense because a large number of abusive combosquatting
domains facilitate malicious network communication for prolonged periods of
time.

Figure~\ref{fig:pbl_malware_time_diff} presents how fast the community comes
across these combosquatting domains. In the cases of domains from the sets
$C_{mal}$ and $C_{pbl}$, we see that most domains are active several months
before they appear in malware traces, or get listed in public black lists. The
only exception is the spam trap that the security vendor is operating, where
more than 50\% of the domain names in the $C_{spa}$ set appear in passive DNS
either a few days before, or on the same day that they appear in the spam trap.
One reasonable explanation for this behavior is that it is an artifact of the
type of abuse (i.e., spam monetization and social engineering) that these
combosquatting domains facilitate.

In order to measure the overall popularity of the domains in the combosquatting
passive DNS ($CP$) dataset over time, in Figure~\ref{fig:overtime_lookups} we
show the DNS lookup volume growth since 2011, according to our PDNS dataset. To
put things into perspective, in the same figure, we plot the lookup volume of
domains in the $C_{abuse}$ set. It is interesting to observe that while the
domains in the $CP$ set have a steady growth over time, the lookup volume of
malicious domain names in the set $C_{abuse}$ appears to be nearly uniform. Even
though we lack a definite explanation of this behavior, our earlier
spam-trap-related results suggest that this almost uniform activity is an
artifact of the type of combosquatting abuse (i.e. related to spam and social
engineering) that the security industry can reliably detect.

Another interesting observation is related to the Alexa popularity of
combosquatting domains. Figure~\ref{fig:alexa_histogram} shows the distributions
of combosquatting domains across the top 1 million Alexa ranks, both for
combosquatting domains that are known to be malicious (present in any of our
abuse datasets) as well as for all of the remaining combosquatting domains.
First, we can observe that, as we move from higher to lower rankings, the
concentration of generic combosquatting domains increases. Even so, the overall
number of combosquatting domains that are present in the top 1 million Alexa
list is limited. In terms of the distribution of malicious combosquatting
domains, there we see the presence of malicious domains across all Alexa ranks,
which suggests that the existing tools for detecting malicious domains are
finding only a small fraction of live attacks, regardless of the overall number
of combosquatting activity in any given bin of Alexa ranking. We should note
that Figure~\ref{fig:alexa_histogram} shows aggregate statistics of 20,000 bins
in the $x$-axis. Therefore, the far left domains are cases of combosquatting
domains that have made it into any of the top 20,000 Alexa ranks.

\subsection{Infrastructure Analysis}
\label{sec:infra}

So far we have examined how the domains in the combosquatting passive DNS
dataset evolved over time. In this section, we turn our attention to the various
DNS and IP properties that the domains in the combosquatting passive and active
DNS dataset exhibit. We see that the hosting infrastructure of malicious
combosquatting domains is concentrated in certain autonomous systems and they
are scattered across numerous different CIDRs---which is different from the
behavior of combosquatting domains in general.

Figure~\ref{fig:cp_ac_infra} shows the distribution of Classless Inter-Domain
Routing (CIDR) networks, Autonomous Systems (AS), and Country Codes (CC) for the
hosting facilities of $CP$ and $CA$ combosquatting domains. As expected, generic
combosquatting activity is spread across the globe with no obvious
concentrations.

We cannot claim the same for the domains in the $C_{abuse}$ set. In
Figure~\ref{fig:mal_infra}, we can see a higher concentration of malicious
combosquatting domains from the $C_{abuse}$ set in a single CIDR and AS. That
is, almost 58\% of the malicious domains are in one CIDR, where only 38\% of all
combosquatting domains live in a single network. The preference that malicious
domains have a single CIDR/AS can be explained in the following two ways. There
are few CIDRs and ASes around the world that will permit the long term hosting
of malicious domains. At the same time, such malicious combosquatting domains
eventually will be remediated, as we saw earlier in this section. This will
practically mean that they will be pointed to a DNS sinkhole or a domain parking
page.

With this behavior in mind, we tried to better understand both the bipartite
graph between the domains in the combosquatting passive and active DNS datasets,
and also in the $C_{abuse}$ set. With Figure~\ref{fig:ips} we observe that
domains in the set $C_{abuse}$ point to hosts that are spread across more
distinct CIDRs than the domains in the $CP$ and $CA$ set. While the rotation on
malicious IP infrastructure might not be a new observation, in the reduced space
of combosquatting domains, this behavior could be used not only as a way to both
track combosquatting domains over time, but also to alert us of potentially new
abusive ones.

\section{Combosquatting in the Wild}
\label{sec:attacks}

So far we have shed light to the combosquatting phenomenon over a period of
almost five years. We have shown the complexity of the combosquatting problem by
studying its lexical, infrastructure, and temporal properties in
Section~\ref{sec:measurements}. This section focuses on how combosquatting
domains are being used in the wild. We study different aspects of combosquatting
abuse, at the time of writing, and show how combosquatting can be used for many
different types of illicit activities.

We show that combosquatting domains are currently being used for a variety of
attacks (e.g. phishing, affiliate abuse, social engineering, trademark abuse).
While we study trademarks spread across different business categories, these
attacks affect almost every category. We manually analyze a set of
combosquatting domains in order to further examine their network behavior and
the countermeasures the adversaries take to evade detection.

\subsection{Exploring \& Labeling Combosquatting Domains}
\label{sub:measuring-abuse}

In order to understand the current status of combosquatting domains and
potential attacks rendered using them, we built an infrastructure of 100
\textit{scriptable} browser instances and used them to crawl 1.3 million
combosquatting domains, which were all part of $CA$ (active DNS dataset). The
1.3 million domains were comprised of 1.13 million initial seed domains (note
that we have slightly more domains than the ones reported in
Table~\ref{table:DS} since we may crawl multiple subdomains per e2LD). On top of
that, we also crawl 200 thousand domains, which included daily registrations of
new combosquatting domains and other domains that switched to unknown NS server
infrastructure (e.g. non-brand protection companies). Our crawlers were tracking
these changes for four weeks and were able to successfully crawl approximately
1.1 million domains.

Due to the sheer size of the collected data and the need of manual verification
by human analysts, we approach the dataset we collected through crawling in
three sequential steps. First, we scan our entire dataset for evidence of
affiliate abuse, i.e., combosquatting domains that redirect users to their
intended destination but add an affiliate identifier while doing so. This check
will result in the scammer earning a commission from the user's
actions~\cite{snyder2015no}. Second, we look in the remainder of the dataset for
phishing pages by identifying login forms (from HTML inspection) and focusing on
the web pages that are ``visually similar'' to the legitimate websites. Finally,
in order to understand the type of abuse that is neither phishing nor affiliate
abuse, we perform a combination of stratified and simple random sampling on our
remaining dataset and manually label 8.7 thousand web pages.

All this effort will yield two important points for our study. First, this will
help the reader get a sense of how combosquatting is currently used in social
engineering and affiliate abuse. Second, we augment the $C_{abuse}$ set of
malicious combosquatting domains that escape the threat feeds we used in our
study. The next paragraph will provide more details about each step and the
discovered abuse.

\paragraph{Affiliate abuse} First, we scan all pages of our crawled corpus
focusing on the ones that, through a series of redirections, navigated our
crawlers to the appropriate authoritative domains. By excluding domains that,
through their WHOIS records and name servers, we identified as clearly belonging
to the legitimate owners of the authoritative domains, we manually investigate
the rest of the redirection chains and identify 2,573 unique domains that were,
for at least one day, involved in affiliate abuse.

\paragraph{Phishing} We scan the HTML code of all the crawled pages that were
neither legitimately owned nor abusing affiliate programs, and identify 40,299
unique domains that contain at least one login form. We then proceed to cluster
these webpages by their visual appearance using a hamming distance on the hashes
produced by a perceptual hashing function, a process which resulted in 7,845
clusters. We then focus on the clusters that contain screenshots that are
similar to the look-and-feel of the targeted brands, so as to remove unrelated
pages that happen to have login forms. Through this process, we identify 174
domains as conducting phishing attacks. Table~\ref{tbl:crawl-phishing} shows the
trademarks that were attacked by four or more combosquatting domains. Even
though this number may appear to be small, these were short-lived \emph{live}
phishing domains that we discovered in the wild targeting the users of our
investigated trademarks.

\paragraph{Other types of abuse} Last, we focus on the top two Alexa domains of
each of the trademark categories (stratified sampling), resulting in the
selection of 221,292 combosquatting domains targeting the selected trademarks.
Using perceptual hashing in the same way as we did for the identification of
phishing pages, we cluster 351 thousand screenshots of websites (note that many
of the 221 thousand combosquatting domains were crawled multiple times due to
infrastructure changes that were deemed suspicious) into 50 thousand clusters.
The trademark responsible for the largest number of clusters (8.3 thousand) was
Amazon which, due to its name, ``attracts'' thousands of combosquatting websites
which are not necessarily related to each other, and thus create clustering
singletons. To label the screenshots, we randomly sample 10\% of the domains of
each affected brand and manually label them, resulting in a manual analysis
effort of 8.7 thousand screenshots.

The labeling was performed by the authors where each one chose among the
following labels: social engineering (surveys, scams such as tech support
scam~\cite{miramirkhani2017tss}, malicious downloads), trademark abuse (websites
capitalizing on the brand of the squatted trademarks), unrelated (seemingly
benign and unrelated websites), and error/under construction. Finally, the
resulting labels are then used to label the entire clusters in which each
sampled screenshot belongs. Table~\ref{tbl:abuse-percentage-overall} shows the
overall abuse of the investigated trademarks by consolidating the results of the
previous two steps, the manual labeling of the stratified random sample and
removing all the authorative domains from the list. Table~\ref{tbl:abusepopular}
shows the types of abuse for each category of trademarks by focusing on the
abuse of its most popular domain (grey cells denote the most popular type of
abuse per trademark category). There we see that while trademark abuse is
usually the most popular type of abuse, the exact breakdown varies across
categories. For example, for both amazon and homedepot, affiliate abuse is the
most popular type of abuse, fueled by the fact that these two services offer
affiliate programs to their users.

\begin{table}[t] \centering
  \begin{tabular}{p{1.75cm} p{1.25cm} p{3.5cm}}
    \textbf{Trademark} & \textbf{\#Phishing} & \textbf{Example}\\
    \hline
    Facebook & 56 & facebook123[.]cf \\
    icloud & 48 & icloudaccountuser[.]com \\
    Amazon & 7 & secure5-amazon[.]com \\
    Google & 8 & drivegoogle[.]ga \\
    PayPal & 8 & paypal-updates[.]ml \\
    Instagram & 7 & wvwinstagram[.]com \\
    Baidu & 4 & baidullhk[.]com \\
    \hline
  \end{tabular}
  \caption{Examples of domains used for phishing, as discovered by our crawling
    infrastructure.}
  \label{tbl:crawl-phishing}
\end{table}

\begin{table}[t]
  \centering
  \begin{threeparttable}
    \begin{tabular}{lllr}
      \hline
      \multirow{2}{*}{\bf Unknown} & \multirow{2}{*}{86.6\%} & Unrelated & 11.23\% \\
                                   & & Suspicious~\tnote{1} & 88.77\% \\
      \hline
      \multirow{4}{*}{\bf Malicious} & \multirow{4}{*}{13.39\%} & Phishing & 0.9\% \\
                                   & & Social Engineering & 13.62\% \\
                                   & & Affiliate Abuse & 15.56\% \\ 
                                   & & Trademark Abuse & 69.9\%  \\
      \hline
    \end{tabular}
    \begin{tablenotes}
    \item[1] Includes under construction, error pages and parking websites.
    \end{tablenotes}
    \caption{Types of combosquatting pages}
    \label{tbl:abuse-percentage-overall}
  \end{threeparttable}
\end{table}

\begin{table}[t]
  \centering
  \begin{adjustbox}{width=\linewidth}
    \begin{tabular}{ p{2.3cm} l r r r r }
      \textbf{Category} & \textbf{Trademark} &  \multicolumn{1}{c}{\textbf{PH}} & \multicolumn{1}{c}{\textbf{AB}} & \multicolumn{1}{c}{\textbf{SE}} & \multicolumn{1}{c}{\textbf{TA}} \\ \hline
      Adult Content & pornhub & 0\% & 5.14\% & 25.73\% & \cellcolor{gray} 69.11\% \\
      Blogging & wordpress & 0\% & 0.06\% & 2.93\% & \cellcolor{gray} 96.96\% \\
      Computers & microsoft & 0.32\% & 11.0\% & 13.68\% & \cellcolor{gray} 74.39\% \\
      E-Shop (Online) & amazon & 0.36\% & \cellcolor{gray} 61.65\% & 1.47\% & 36.50\% \\
      Financial & paypal & 6.29\% & 0.78\% & \cellcolor{gray} 55.11\% & 37.79\% \\
      Radio \& TV & netflix & 2.29\% & 5.74\% & 19.54\% & \cellcolor{gray} 72.41\% \\
      E-Learning & wikipedia & 0\% & 0\% & 32.58\% & \cellcolor{gray} 67.14\% \\
      Lifestyle & diply & 0\% & 0\% & 1.6\% & \cellcolor{gray} 98.4\% \\
      News & reddit & 1.49\% &0\% & 1.49\% & \cellcolor{gray} 97.01\% \\
      Couriers & fedex & 0\% & 3.12\% & 25\% & \cellcolor{gray} 71.87\% \\
      E-Shop (C2C) & craigslist & 0\% & 0\% & 31.10\% & \cellcolor{gray} 68.89\% \\
      Photography & pinterest & 0\% & 0\% & 5.76\% &\cellcolor{gray} 94.23\% \\
      E-Shop (Physical)  & homedepot & 0\% & \cellcolor{gray} 72.5\% & 2.5\% & 25\% \\
      Search Engines & google & 0.32\% & 3.58\% & 23.49\% & \cellcolor{gray} 72.32\% \\
      File Sharing & dropbox & 2.7\% & 16.21\% &\cellcolor{gray}  51.35\% & 29.72\% \\
      Social Networks & facebook & 5.24\% & 6.18\% & 18.74\% & \cellcolor{gray} 69.82\% \\
      Software \& Web & popads & 0\% & 0\% & 0\% & \cellcolor{gray} 100\% \\
      Streaming & youtube & 0\% & 2.02\% & 14.5\% & \cellcolor{gray} 83.47\% \\
      Telecom & xfinity & 2.85\% & 14.28\% & 11.42\% & \cellcolor{gray} 71.42\% \\
      Travel & airbnb & 0\% & 4.04\% & 1\% & \cellcolor{gray} 94.95\% \\
      \hline
    \end{tabular}
  \end{adjustbox}
  \caption{
    Types of combosquatting abuse for the most popular investigated domain
    within each trademark category (PH = phishing, AB = affiliate abuse, SE =
    social engineering, TA = trademark abuse).
  } 
  \label{tbl:abusepopular}
\end{table}

\subsection{Case Studies} 
\label{sub:case-study}

On October 30th of 2016, we crawled 505 combosquatting domain names that were
hosted on the same infrastructure. That is, the domain names were pointing to
the same set of IP addresses on that day according to the active DNS dataset. To
better understand how adversaries take advantage of combosquatting domains, we
set up a headless crawling engine based on the Python \textit{requests} module,
that collects Layer 7 (in the OSI stack) information. Our experimental setup had
two phases: first we crawled the domains using the default configuration of the
module and then we repeated the process specifying a Chrome \textit{User-Agent}.
By comparing crawling results from the two phases, we were able to identify the
presence of evasive behavior against our crawlers based on factors like HTTP
headers, client's IP address and cookies' presence.

\paragraph{Redirection Games}
Most of the domains were associated with a form of redirection, either to a
parking page, or to an abuse-related website. A set of 114 domains were
performing at least one redirection irrespective of the User-Agent HTTP header.
When the User-Agent was not set, 28 domains did not redirect and presented a
parking page. This set grew to 127 when User-Agent headers were used.
Redirection to the parking page was performed via a child label for the same
domain name, following the same naming convention: the child label starts with
{\tt ww} followed by a number (i.e. starbucksben[.]com redirects to
ww1.starbucksben[.]com).

\begin{figure}[t]
  \begin{center}
    \includegraphics[scale=0.6]{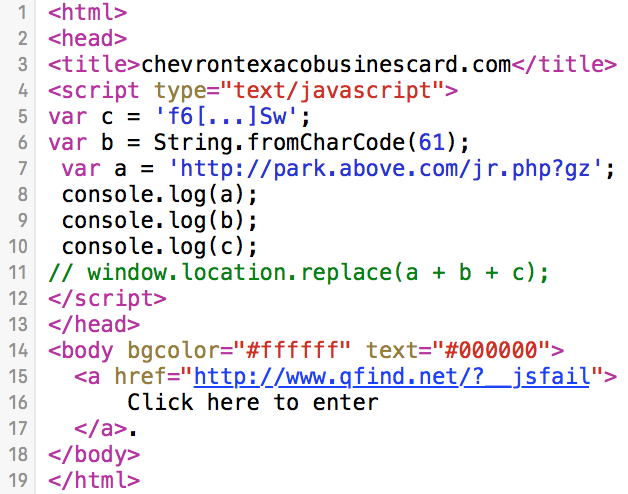}
    \caption{The JavaScript redirection performed by some combosquatting domain
      names. This example is the result of visiting
      chevrontexacobusinescard[.]com. Line 5 had a 1,838 characters long
      string.}
    \label{fig:js_redirect}
  \end{center}
\end{figure}

Moreover, there was a set of 53 domains that was performing HTTP redirection
without User-Agent, but JavaScript redirection when the User-Agent was set. In
the latter case, the HTTP response contained highly obfuscated JavaScript code
similar to the one in Figure~\ref{fig:js_redirect}.

\paragraph{Malware Drops}
One interesting example that shows how adversaries are hiding the behavior of a
domain from automated systems and crawlers, is http://zillowhomesforsale[.]com.
When no User-Agent is present, the domain always redirected to
http://ww1.zillowhomesforsale[.]com/, which served us with a parking template.
When the User-Agent was set, the redirection would be to either the
aforementioned URL or to a completely different domain (i.e.
http://rtbtracking[.]com/click?data=Mm[...]Q2\&id=8c[...]d3), based on a
probabilistic algorithm.

After we identified the attempt of the domains to hide their real behavior, we
tried to extract further information. We setup two Virtual Machines (VMs) on a
MacBook Pro running Mac OS 10.11.6 and Avast Mac Security 2015 Version 11.18
(46914) with Virus definitions version 16103000. The first VM was an Ubuntu
14.04.1 and the second a Mac OS 10.11.6. We started manually browsing to the
domain names mentioned earlier and we identified several instances of malicious
websites and URLs we were redirected to.

For example, zillowhomesforsale[.]com this time redirected us to
http://www.searchnet[.]com/Search/Loading?v=5 which was blocked by Avast and
classified as \textit{RedirMe-inf [Trj]}, a well known trojan
(http://malwarefixes.com/threats/htmlredirme-inf-trj/). Similarly, when we
browsed to youtubezeneletoltes[.]net we came across an automatic downloader of a
disk image file named ``FlashPlayer.dmg''. It contained a binary that we
submitted to VirusTotal for analysis. The results pointed to malware, since
15/54 Antivirus reports were suggesting some type of Trojan or Adware
(\url{http://bit.ly/2ffwyW1}).

Some of the combosquatting domains that we experimented with would redirect us
to an authoritative website (not necessarily the one they were abusing), after
appending an affiliate identifier in the URL, essentially conducting affiliate
abuse. For instance, visiting jcpenneyoulet[.]com lands on
http://www.jcpenney[.]com/?cm\_mmc=google\%20non-[...] and visiting
toysrusuk[.]com yields http://www.target[.]com/?clkid=4738[...]. Interestingly,
after visiting one of the websites a cookie would be set on the user's browser.
If the user attempted to visit another website (from the same set of domains),
she would find herself on a parking page~\cite{parking_ndss2015}. After clearing
cookies and repeating the process, the domain would reveal its true nature.

\paragraph{Social Engineering and Phishing}
Another type of abuse we identified was related to social engineering and
phishing types of attacks. After visiting some domains like
\underline{staples}easeyrebates[.]com, we were redirected to
http://viewcustomer[.]com/s3/p10/index-20up-p10-cnf-t1-p4.php?tracker=wait.loading-links.com\&keyword=staples1[...].
The landing page presented us with a survey for Staples that would supposedly
reward us with a gift after completing it, clearly not related to the Staples
business in any way. These surveys are meant to collect as much PII as possible
from users and subscribe them to potentially paid
services~\cite{clark2013there}.

\section{Discussion}

In Sections~\ref{sec:background} through~\ref{sec:attacks}, we presented the
intuitions behind combosquatting domains, how they are different than other
types of domain squatting, and quantified their current level of abuse.
Specifically, we found that combosquatting, despite its relative obscurity, is
more popular than typosquatting (Section~\ref{sec:combo-vs-typo}). Further, we
found that combosquatters carefully crafted their domain names to account for
the businesses to which the abused trademarks belong to
(Section~\ref{sec:lexical}). By cross-referencing our list of combosquatting
domains with popular blacklists, we observed that most domains are active for
several months before appearing on these lists (Section~\ref{sec:temporal}),
suggesting the presence of blind spots in the tools used by the security
industry. We identified that a few ASes were responsible for the long-term
hosting of malicious combosquatting domains (Section~\ref{sec:infra}) and
witnessed how both common botnets and targeted APTs utilize combosquatting
domains to benefit from trademark recognition and remain hidden in plain sight.
Finally, by actively crawling 1.3 million combosquatting domains and labeling
the results, we witnessed live phishing domains and the abuse of trademarks
across all studied business categories (Section~\ref{sub:measuring-abuse}).

Given the magnitude of the combosquatting problem, in this section, we discuss
what can be done in terms of countermeasures against combosquatting
from the viewpoint of different actors in the domain name ecosystem. \\

\noindent\textbf{Registrants.} A defense that is commonly proposed against
traditional types of domain squatting are \emph{defensive registrations}. In
defensive registrations, companies can proactively register domains that are
likely to be abused (e.g. Microsoft owns wwwmicrosoft[.]com which redirects
users to microsoft[.]com) before miscreants have a chance of registering them.
Combosquatting, however, is unique in the sense that it lacks a generative model
(discussed in Section~\ref{sec:combo-vs-typo}). As such, even for companies that
can afford a large number of defensive registrations, there is no single
algorithm (like the typo models of Wang et al.~\cite{wang2006strider}) that
could be used to generate a list of combosquatting domains. Therefore, the
burden of protecting against combosquatting domains cannot rest on registrants.

At the same time, we consider it of crucial importance that tradermark owners
stop utilizing the practice of registering benign combosquatting domains for
their business. For example, the domain paypal-prepaid[.]com belongs to PayPal
and advertises the ability to use PayPal to obtain prepaid debit cards. By using
these types of domains, companies are indirectly training users that domains
that contain their trademark are legitimate, making it harder for every day
users to detect the malicious ones (which, as discussed in
Section~\ref{sec:combo-vs-typo}, will also have TLS certificates). Instead,
trademark owners can use filepaths (e.g. www.paypal[.]com/prepaid), subdomains
(e.g. prepaid.paypal[.]com) or even TLDs (e.g. prepaid[.]paypal) to advertise
their products without the risks associated with the registration of
combosquatting domains. \\

\noindent\textbf{Registrars.} Registrars are in the unique position to know
which domains users are trying to register before they actually register them.
Therefore, we argue that registrars could add extra logic in their
fraud-detection systems to flag domains that contain popular trademarks
(following a process similar to ours). For each flagged domain, the registrar
can either request more information from the users who attempt to register them,
or follow up on those domains to ensure that they are not used for malicious
purposes. Even though there will always be registrars who do not wish to
implement such countermeasures and who turn a blind eye to abuse, over time,
these registrars and all the domains registered through them could be treated by
domain-intelligence systems as ``suspicious.'' This unwanted labeling will
translate to loss of income forcing registrars to either adopt
fraud-detection systems, or risk further loss of business. \\

\noindent\textbf{Third parties.} Next to registrars, there exist a wide array
of systems~\cite{pleiades, kopis, notos, hao2016predator, hao2013understanding,
  hao2011monitoring} which analyze newly registered domains in an attempt to
discover abusive ones before they are weaponized. Similar to the extra step for
registrars, we argue that searching for the presence of popular trademarks in
newly registered domains can be an extra source of signal that can be exploited
to identify malicious registrations.

\section{Related Work}
\label{sec:related-work}

\noindent\textbf{DNS Abuse.}
Weimer et al.~\cite{weimer05} proposed collecting passive DNS data for security
analysis. Since then, researchers have used passive DNS data to build domain
name reputation systems using statistical modeling methods to detect abuse on
the
Internet~\cite{notos,exposure,kopis,pleiades,ma2009beyond,6903614,7266868,5958207}.
More recently, Lever et al.~\cite{lever2016domain} used passive DNS to identify
potential domain ownership changes. Hao et al.~\cite{hao2016predator} uses only
registration features to build domain reputation system. Liu et
al.~\cite{liu2016all} revealed that dangling DNS records pointing to invalid
resources can be easily manipulated for domain highjacking. Chen et
al.~\cite{sinkanalysis} used passive DNS data to estimate the financial abuse of
advertising ecosystem by a large botnet. \\

\noindent\textbf{Squatting Abuse.}
Several studies have focused on domain squatting in general. Jakobsson et
al.~\cite{jakobsson2007instills, jakobsson2007human} proposed techniques for
identifying typosquatting and discovered that websites in categories with higher
PPC ad prices face more typosquatting registrations. Wang et
al.~\cite{wang2006strider} proposed models for the generation of typosquatting
domains from authoritative ones. Agten et al.~\cite{agten2015seven} studied
typosquatting using crawled data over a period of seven months finding, among
others, that few trademark owners protect themselves by defensively registering
typosquatting domains. In addition to typosquatting, Nikiforakis et
al.~\cite{nikiforakis2013bitsquatting} quantify the extent to which attackers
are leveraging bitsquatting~\cite{dinaburg:20011}, where random bit-errors
occurring in the memory of commodity hardware can redirect Internet traffic to
attacker-controlled domains. Their experiments show that new bitsquatting
domains are registered daily and monetized through ads, affiliate programs and
even malware installations. The authors later performed a measurement of the
so-called ``soundsquatting'', where attackers abuse homophones to attract users
and confuse text-to-speech systems~\cite{nikiforakis2014soundsquatting}.

The only work on combosquatting other than this paper is a brief 2008 industry
whitepaper~\cite{combosquatting2008}. Starting with 30 trademarks and up to 50
generic keywords the authors constructed possible combosquatting domains and
then attempted to get traffic data for the 500 domains that were registered. The
authors found that most sites were filled with ads, thereby abusing the
popularity of trademarks and diluting their revenue. Motivated by the findings
of that nine-year-old whitepaper, we performed the experiments described in this
paper finding millions of combosquatting domains and analyzing registration and
abuse trends over almost six years.

\section{Conclusion}
\label{section:conclusions}

In this paper, we study a type of domain squatting termed ``combosquatting,''
which has yet to be extensively studied by the security community. By
registering domains that include popular trademarks (e.g.,
\url{paypal-members[.]com}), attackers are able to capitalize on a trademark's
recognition to perform social engineering, phishing, affiliate abuse, trademark
abuse, and even targeted attacks. We performed the first large-scale, empirical
study of combosquatting using 468 billion DNS records from both active and
passive DNS datasets, which were collected over an almost six year time period.
Lexical analysis of combosquatting domains revealed that, while there is an
almost infinite pool of potential combosquatting domains, most instances add
only a single token to the original combosquatted domain. Furthermore, the
chosen tokens were often specifically targeted to a particular business
category. These results can help brands limit the potential search space for
combosquatting domains. Additionally, our results show that most combosquatting
domains were not remediated for extended periods of times---up to 1,000 days in
many cases. Furthermore, many instances of combosquatting abuse were seen active
significantly before they were discovered by public blacklists or malware feeds.
Consequently, our findings suggests that current protections do not do a good
job at addressing the threat of combosquatting. This is particularly concerning
because our results also show that combosquatting is becoming more prevalent
year over year. Lastly, we found numerous instances of combosquatting abuse in
the real world by crawling 1.3 million combosquatting domains and manually
analyzing the results. Based on our findings we discuss the role of different
parties in the domain name ecosystem and how each party can help tackle the
overall combosquatting problem. Ultimately, our results suggest that
combosquatting is a real and growing threat, and the security community needs to
develop better protections to defend against it.

\begin{acks}
  
  The authors would like to thank the anonymous reviewers for their valuable
  comments and helpful suggestions.

  This material is based upon work supported in part by the
  \grantsponsor{DoC}{US Department of Commerce}{} under Grant
  No.:~\grantnum{DoC}{2106DEK} and~\grantnum{DoC}{2106DZD}; the
  \grantsponsor{NSF}{National Science Foundation (NSF)}{} under Grant
  No.:~\grantnum{NSF}{2106DGX}, \grantnum{NSF}{CNS-1617902},
  \grantnum{NSF}{CNS-1617593}, and \grantnum{NSF}{CNS-1735396}; the
  \grantsponsor{DARPA}{Air Force Research Laboratory/Defense Advanced Research
    Projects Agency}{} under Grant No.:~\grantnum{DARPA}{2106DTX}; and the
  \grantsponsor{OoN}{Office of Naval Research (ONR)}{} under Grant
  No.:~\grantnum{OoN}{N00014-16-1-2264}.

  Any opinions, findings, and conclusions or recommendations expressed in this
  material are those of the authors and do not necessarily reflect the views of
  the US Department of Commerce, National Science Foundation, Air Force Research
  Laboratory, Defense Advanced Research Projects Agency, nor Office of Naval
  Research.

\end{acks}
\balance
\bibliographystyle{ACM-Reference-Format}
\bibliography{sigproc,rfc,apt}


\begin{thebibliography}{00}


\ifx \showCODEN    \undefined \def \showCODEN     #1{\unskip}     \fi
\ifx \showDOI      \undefined \def \showDOI       #1{#1}\fi
\ifx \showISBNx    \undefined \def \showISBNx     #1{\unskip}     \fi
\ifx \showISBNxiii \undefined \def \showISBNxiii  #1{\unskip}     \fi
\ifx \showISSN     \undefined \def \showISSN      #1{\unskip}     \fi
\ifx \showLCCN     \undefined \def \showLCCN      #1{\unskip}     \fi
\ifx \shownote     \undefined \def \shownote      #1{#1}          \fi
\ifx \showarticletitle \undefined \def \showarticletitle #1{#1}   \fi
\ifx \showURL      \undefined \def \showURL       {\relax}        \fi
\providecommand\bibfield[2]{#2}
\providecommand\bibinfo[2]{#2}
\providecommand\natexlab[1]{#1}
\providecommand\showeprint[2][]{arXiv:#2}

\bibitem[\protect\citeauthoryear{??}{com}{2008}]%
        {combosquatting2008}
 \bibinfo{year}{2008}\natexlab{}.
\newblock \showarticletitle{{Combosquatting: The Business of Cybersquatting}}.
  In \bibinfo{booktitle}{{\em {FairWinds Partners, LLC }}}.
\newblock


\bibitem[\protect\citeauthoryear{??}{pbl}{2015}]%
        {pbl_driveby}
 \bibinfo{year}{2015}\natexlab{}.
\newblock \bibinfo{title}{{Domain Blacklist: driveby}}.
\newblock
  \bibinfo{howpublished}{\url{http://www.blade-defender.org/eval-lab/}}.
  (\bibinfo{year}{2015}).
\newblock


\bibitem[\protect\citeauthoryear{??}{pbl}{2016a}]%
        {pbl_abusech}
 \bibinfo{year}{2016}\natexlab{a}.
\newblock \bibinfo{title}{{Domain Blacklist: abuse.ch}}.
\newblock \bibinfo{howpublished}{\url{http://www.abuse.ch/}}.
  (\bibinfo{year}{2016}).
\newblock


\bibitem[\protect\citeauthoryear{??}{pbl}{2016b}]%
        {pbl_malc0de}
 \bibinfo{year}{2016}\natexlab{b}.
\newblock \bibinfo{title}{{Domain Blacklist: Blackhole {DNS}}}.
\newblock
  \bibinfo{howpublished}{\url{http://www.malwaredomains.com/wordpress/?page_id=6}}.
    (\bibinfo{year}{2016}).
\newblock


\bibitem[\protect\citeauthoryear{??}{pbl}{2016c}]%
        {pbl_hphosts}
 \bibinfo{year}{2016}\natexlab{c}.
\newblock \bibinfo{title}{{Domain Blacklist: hphosts}}.
\newblock \bibinfo{howpublished}{\url{http://hosts-file.net/?s=Download}}.
  (\bibinfo{year}{2016}).
\newblock


\bibitem[\protect\citeauthoryear{??}{pbl}{2016d}]%
        {pbl_itmate}
 \bibinfo{year}{2016}\natexlab{d}.
\newblock \bibinfo{title}{{Domain Blacklist: itmate}}.
\newblock \bibinfo{howpublished}{\url{http://vurl.mysteryfcm.co.uk/}}.
  (\bibinfo{year}{2016}).
\newblock


\bibitem[\protect\citeauthoryear{??}{pbl}{2016e}]%
        {pbl_sagadc}
 \bibinfo{year}{2016}\natexlab{e}.
\newblock \bibinfo{title}{{Domain Blacklist: sagadc}}.
\newblock \bibinfo{howpublished}{\url{http://dns-bh.sagadc.org/}}.
  (\bibinfo{year}{2016}).
\newblock


\bibitem[\protect\citeauthoryear{??}{pbl}{2016f}]%
        {pbl_sans}
 \bibinfo{year}{2016}\natexlab{f}.
\newblock \bibinfo{title}{{Domain Blacklist: SANS}}.
\newblock
  \bibinfo{howpublished}{\url{https://isc.sans.edu/suspicious_domains.html}}.
  (\bibinfo{year}{2016}).
\newblock


\bibitem[\protect\citeauthoryear{??}{pbl}{2016g}]%
        {pbl_malwaredomainlist}
 \bibinfo{year}{2016}\natexlab{g}.
\newblock \bibinfo{title}{{Malware Domain List}}.
\newblock
  \bibinfo{howpublished}{\url{http://www.malwaredomainlist.com/forums/index.php?topic=3270.0}}.
    (\bibinfo{year}{2016}).
\newblock


\bibitem[\protect\citeauthoryear{??}{cer}{2017}]%
        {certificate-transparency}
 \bibinfo{year}{2017}\natexlab{}.
\newblock \bibinfo{title}{Certificate Transparency}.
\newblock
  \bibinfo{howpublished}{\url{https://www.certificate-transparency.org}}.
  (\bibinfo{year}{2017}).
\newblock


\bibitem[\protect\citeauthoryear{Aas}{Aas}{2015}]%
        {cas-role-malware}
\bibfield{author}{\bibinfo{person}{Josh Aas}.} \bibinfo{year}{2015}\natexlab{}.
\newblock \bibinfo{title}{{Let's Encrypt: The CA's Role in Fighting Phishing
  and Malware}}.
\newblock
  \bibinfo{howpublished}{\url{https://letsencrypt.org/2015/10/29/phishing-and-malware.html}}.
    (\bibinfo{year}{2015}).
\newblock


\bibitem[\protect\citeauthoryear{ACPA}{ACPA}{1999}]%
        {acpa}
ACPA \bibinfo{year}{1999}\natexlab{}.
\newblock \bibinfo{title}{{Anticybersquatting Consumer Protection Act (ACPA)}}.
\newblock \bibinfo{howpublished}{\url{http://www.patents.com/acpa.htm}}.
  (\bibinfo{date}{November} \bibinfo{year}{1999}).
\newblock


\bibitem[\protect\citeauthoryear{{Agten, Pieter and Joosen, Wouter and
  Piessens, Frank and Nikiforakis, Nick}}{{Agten, Pieter and Joosen, Wouter and
  Piessens, Frank and Nikiforakis, Nick}}{2015}]%
        {agten2015seven}
\bibfield{author}{\bibinfo{person}{{Agten, Pieter and Joosen, Wouter and
  Piessens, Frank and Nikiforakis, Nick}}.} \bibinfo{year}{2015}\natexlab{}.
\newblock \showarticletitle{{Seven months' worth of mistakes: A longitudinal
  study of typosquatting abuse}}. In \bibinfo{booktitle}{{\em {Proceedings of
  the 22nd Network and Distributed System Security Symposium (NDSS 2015)}}}.
  {Internet Society}.
\newblock


\bibitem[\protect\citeauthoryear{Alexa}{Alexa}{2016}]%
        {alexa}
\bibfield{author}{\bibinfo{person}{Alexa}.} \bibinfo{year}{2016}\natexlab{}.
\newblock \bibinfo{title}{{The Web Information Company}}.
\newblock \bibinfo{howpublished}{\url{http://www.alexa.com/}}.
  (\bibinfo{year}{2016}).
\newblock


\bibitem[\protect\citeauthoryear{{AllSlang}}{{AllSlang}}{2016a}]%
        {noslang}
\bibfield{author}{\bibinfo{person}{{AllSlang}}.}
  \bibinfo{year}{2016}\natexlab{a}.
\newblock \bibinfo{title}{{Slang Dictionary - Text Slang \& Internet Slang
  Words}}.
\newblock \bibinfo{howpublished}{\url{http://www.noslang.com/dictionary/}}.
  (\bibinfo{year}{2016}).
\newblock


\bibitem[\protect\citeauthoryear{{AllSlang}}{{AllSlang}}{2016b}]%
        {noswearing}
\bibfield{author}{\bibinfo{person}{{AllSlang}}.}
  \bibinfo{year}{2016}\natexlab{b}.
\newblock \bibinfo{title}{{Swear Word List \& Curse Filter}}.
\newblock \bibinfo{howpublished}{\url{http://www.noswearing.com/dictionary}}.
  (\bibinfo{year}{2016}).
\newblock


\bibitem[\protect\citeauthoryear{{Anton Cherepanov}}{{Anton
  Cherepanov}}{2014}]%
        {roaming_tiger}
\bibfield{author}{\bibinfo{person}{{Anton Cherepanov}}.}
  \bibinfo{year}{2014}\natexlab{}.
\newblock \bibinfo{title}{{ScanBox framework --- who's affected, and who's
  using it?}}
\newblock
  \bibinfo{howpublished}{\url{http://2014.zeronights.org/assets/files/slides/roaming_tiger_zeronights_2014.pdf}}.
    (\bibinfo{date}{{July}} \bibinfo{year}{2014}).
\newblock


\bibitem[\protect\citeauthoryear{Antonakakis, Perdisci, Dagon, Lee, and
  Feamster}{Antonakakis et~al\mbox{.}}{2010}]%
        {notos}
\bibfield{author}{\bibinfo{person}{Manos Antonakakis}, \bibinfo{person}{Roberto
  Perdisci}, \bibinfo{person}{David Dagon}, \bibinfo{person}{Wenke Lee}, {and}
  \bibinfo{person}{Nick Feamster}.} \bibinfo{year}{2010}\natexlab{}.
\newblock \showarticletitle{Building a Dynamic Reputation System for {DNS}}. In
  \bibinfo{booktitle}{{\em { the Proceedings of 19th USENIX Security Symposium
  (USENIX Security '10)}}}.
\newblock


\bibitem[\protect\citeauthoryear{Antonakakis, Perdisci, Lee, Vasiloglou, and
  Dagon}{Antonakakis et~al\mbox{.}}{2011}]%
        {kopis}
\bibfield{author}{\bibinfo{person}{Manos Antonakakis}, \bibinfo{person}{Roberto
  Perdisci}, \bibinfo{person}{Wenke Lee}, \bibinfo{person}{Nikolaos
  Vasiloglou}, {and} \bibinfo{person}{David Dagon}.}
  \bibinfo{year}{2011}\natexlab{}.
\newblock \showarticletitle{Detecting Malware Domains in the Upper {DNS}
  Hierarchy}. In \bibinfo{booktitle}{{\em { the Proceedings of 20th USENIX
  Security Symposium (USENIX Security '11)}}}.
\newblock


\bibitem[\protect\citeauthoryear{Antonakakis, Perdisci, Nadji, Vasiloglou,
  Abu-Nimeh, Lee, and Dagon}{Antonakakis et~al\mbox{.}}{2012}]%
        {pleiades}
\bibfield{author}{\bibinfo{person}{Manos Antonakakis}, \bibinfo{person}{Roberto
  Perdisci}, \bibinfo{person}{Yacin Nadji}, \bibinfo{person}{Nikolaos
  Vasiloglou}, \bibinfo{person}{Saeed Abu-Nimeh}, \bibinfo{person}{Wenke Lee},
  {and} \bibinfo{person}{David Dagon}.} \bibinfo{year}{2012}\natexlab{}.
\newblock \showarticletitle{{From Throw-Away Traffic to Bots: Detecting the
  Rise of DGA-Based Malware}}. In \bibinfo{booktitle}{{\em { the Proceedings of
  21th USENIX Security Symposium (USENIX Security '12)}}}.
\newblock


\bibitem[\protect\citeauthoryear{{Asert}}{{Asert}}{2014}]%
        {ixeshe1}
\bibfield{author}{\bibinfo{person}{{Asert}}.} \bibinfo{year}{2014}\natexlab{}.
\newblock \bibinfo{title}{{Illuminating the Etumbot APT Backdoor}}.
\newblock
  \bibinfo{howpublished}{\url{https://github.com/kbandla/APTnotes/blob/master/2014/ASERT-Threat-Intelligence-Brief-2014-07-Illuminating-Etumbot-APT.pdf}}.
    (\bibinfo{date}{{June}} \bibinfo{year}{2014}).
\newblock


\bibitem[\protect\citeauthoryear{{Asert}}{{Asert}}{2016a}]%
        {four_element_sword}
\bibfield{author}{\bibinfo{person}{{Asert}}.} \bibinfo{year}{2016}\natexlab{a}.
\newblock \bibinfo{title}{{The Four Element Sword Engagement}}.
\newblock
  \bibinfo{howpublished}{\url{https://www.arbornetworks.com/blog/asert/four-element-sword-engagement/}}.
    (\bibinfo{date}{{April}} \bibinfo{year}{2016}).
\newblock


\bibitem[\protect\citeauthoryear{{Asert}}{{Asert}}{2016b}]%
        {trochilus}
\bibfield{author}{\bibinfo{person}{{Asert}}.} \bibinfo{year}{2016}\natexlab{b}.
\newblock \bibinfo{title}{{Uncovering the Seven Pointed Dagger Discovery of the
  Trochilus RAT and Other Targeted Threats}}.
\newblock \bibinfo{howpublished}{\url{https://goo.gl/zMbqpA}}.
  (\bibinfo{date}{{January}} \bibinfo{year}{2016}).
\newblock


\bibitem[\protect\citeauthoryear{{Athanasios Kountouras and Panagiotis Kintis
  and Chaz Lever and Yizheng Chen and Yacin Nadji and David Dagon and Manos
  Antonakakis and Rodney Joffe}}{{Athanasios Kountouras and Panagiotis Kintis
  and Chaz Lever and Yizheng Chen and Yacin Nadji and David Dagon and Manos
  Antonakakis and Rodney Joffe}}{2016}]%
        {activedns}
\bibfield{author}{\bibinfo{person}{{Athanasios Kountouras and Panagiotis Kintis
  and Chaz Lever and Yizheng Chen and Yacin Nadji and David Dagon and Manos
  Antonakakis and Rodney Joffe}}.} \bibinfo{year}{2016}\natexlab{}.
\newblock \showarticletitle{{Enabling Network Security Through Active {DNS}
  Datasets}}. In \bibinfo{booktitle}{{\em {Research in Attacks, Intrusions, and
  Defenses - 19th International Symposium, {RAID} 2016, Paris, France,
  September 19-21, 2016, Proceedings}}}. \bibinfo{pages}{188--208}.
\newblock
\showDOI{%
\url{https://doi.org/10.1007/978-3-319-45719-2_9}}


\bibitem[\protect\citeauthoryear{Bilge, Kirda, Kruegel, and Balduzzi}{Bilge
  et~al\mbox{.}}{2011}]%
        {exposure}
\bibfield{author}{\bibinfo{person}{Leyla Bilge}, \bibinfo{person}{Engin Kirda},
  \bibinfo{person}{Christopher Kruegel}, {and} \bibinfo{person}{Marco
  Balduzzi}.} \bibinfo{year}{2011}\natexlab{}.
\newblock \showarticletitle{{EXPOSURE}: Finding Malicious Domains Using Passive
  DNS Analysis}. In \bibinfo{booktitle}{{\em Proceedings of NDSS}}.
\newblock


\bibitem[\protect\citeauthoryear{{Bitdefender}}{{Bitdefender}}{2013}]%
        {miniduke}
\bibfield{author}{\bibinfo{person}{{Bitdefender}}.}
  \bibinfo{year}{2013}\natexlab{}.
\newblock \bibinfo{title}{{A Closer Look at MiniDuke}}.
\newblock
  \bibinfo{howpublished}{\url{https://labs.bitdefender.com/wp-content/uploads/downloads/2013/04/MiniDuke_Paper_Final.pdf}}.
    (\bibinfo{date}{{May}} \bibinfo{year}{2013}).
\newblock


\bibitem[\protect\citeauthoryear{{CHECK POINT SOFTWARE TECHNOLOGIES}}{{CHECK
  POINT SOFTWARE TECHNOLOGIES}}{2015}]%
        {rocket_kitten}
\bibfield{author}{\bibinfo{person}{{CHECK POINT SOFTWARE TECHNOLOGIES}}.}
  \bibinfo{year}{2015}\natexlab{}.
\newblock \bibinfo{title}{{ROCKET KIT TEN: A CAMPAIGN WITH 9 LIVES}}.
\newblock
  \bibinfo{howpublished}{\url{http://blog.checkpoint.com/wp-content/uploads/2015/11/rocket-kitten-report.pdf}}.
    (\bibinfo{date}{{November}} \bibinfo{year}{2015}).
\newblock


\bibitem[\protect\citeauthoryear{{Chen, Yizheng and Kintis, Panagiotis and
  Antonakakis, Manos and Nadji, Yacin and Dagon, David and Lee, Wenke and
  Farrell, Michael}}{{Chen, Yizheng and Kintis, Panagiotis and Antonakakis,
  Manos and Nadji, Yacin and Dagon, David and Lee, Wenke and Farrell,
  Michael}}{2016}]%
        {sinkanalysis}
\bibfield{author}{\bibinfo{person}{{Chen, Yizheng and Kintis, Panagiotis and
  Antonakakis, Manos and Nadji, Yacin and Dagon, David and Lee, Wenke and
  Farrell, Michael}}.} \bibinfo{year}{2016}\natexlab{}.
\newblock \showarticletitle{{Financial Lower Bounds of Online Advertising
  Abuse}}. In \bibinfo{booktitle}{{\em {Proceedings of the 13th International
  Conference on Detection of Intrusions and Malware, and Vulnerability
  Assessment-Volume 9721}}}. {Springer-Verlag New York, Inc.},
  \bibinfo{pages}{231--254}.
\newblock


\bibitem[\protect\citeauthoryear{Clark and McCoy}{Clark and McCoy}{2013}]%
        {clark2013there}
\bibfield{author}{\bibinfo{person}{Jason~W Clark} {and} \bibinfo{person}{Damon
  McCoy}.} \bibinfo{year}{2013}\natexlab{}.
\newblock \showarticletitle{{There Are No Free iPads: An Analysis of Survey
  Scams as a Business.}}. In \bibinfo{booktitle}{{\em LEET}}.
\newblock


\bibitem[\protect\citeauthoryear{{Cylance}}{{Cylance}}{2016}]%
        {dust_storm}
\bibfield{author}{\bibinfo{person}{{Cylance}}.}
  \bibinfo{year}{2016}\natexlab{}.
\newblock \bibinfo{title}{{OPERATION DUST STORM}}.
\newblock
  \bibinfo{howpublished}{\url{https://www.cylance.com/hubfs/2015_cylance_website/assets/operation-dust-storm/Op_Dust_Storm_Report.pdf?t=1477417126448}}.
    (\bibinfo{date}{{February}} \bibinfo{year}{2016}).
\newblock


\bibitem[\protect\citeauthoryear{Dinaburg}{Dinaburg}{2011}]%
        {dinaburg:20011}
\bibfield{author}{\bibinfo{person}{Artem Dinaburg}.}
  \bibinfo{year}{2011}\natexlab{}.
\newblock \showarticletitle{{Bitsquatting: DNS Hijacking without
  Exploitation}}. In \bibinfo{booktitle}{{\em Proceedings of BlackHat
  Security}}.
\newblock


\bibitem[\protect\citeauthoryear{{dmoz}}{{dmoz}}{2016}]%
        {dmoz}
\bibfield{author}{\bibinfo{person}{{dmoz}}.} \bibinfo{year}{2016}\natexlab{}.
\newblock \bibinfo{title}{{DMOZ - the Open Directory Project}}.
\newblock \bibinfo{howpublished}{\url{http://www.dmoz.org}}.
  (\bibinfo{year}{2016}).
\newblock


\bibitem[\protect\citeauthoryear{{Edelman, Benjamin}}{{Edelman,
  Benjamin}}{2003}]%
        {edelman2003}
\bibfield{author}{\bibinfo{person}{{Edelman, Benjamin}}.}
  \bibinfo{year}{2003}\natexlab{}.
\newblock \showarticletitle{{Large-scale registration of domains with
  typographical errors}}.
\newblock \bibinfo{journal}{{\em {Harvard University}\/}}
  (\bibinfo{year}{2003}).
\newblock


\bibitem[\protect\citeauthoryear{{Fidelis Threat Research Team}}{{Fidelis
  Threat Research Team}}{2016}]%
        {turbo_twist}
\bibfield{author}{\bibinfo{person}{{Fidelis Threat Research Team}}.}
  \bibinfo{year}{2016}\natexlab{}.
\newblock \bibinfo{title}{{Turbo Twist: Two 64-bit Derusbi Strains Converge}}.
\newblock
  \bibinfo{howpublished}{\url{http://www.threatgeek.com/2016/05/turbo-twist-two-64-bit-derusbi-strains-converge.html}}.
    (\bibinfo{date}{{May}} \bibinfo{year}{2016}).
\newblock


\bibitem[\protect\citeauthoryear{{FireEye}}{{FireEye}}{2013a}]%
        {saffron_rose}
\bibfield{author}{\bibinfo{person}{{FireEye}}.}
  \bibinfo{year}{2013}\natexlab{a}.
\newblock \bibinfo{title}{{OPERATION SAFFRON ROSE}}.
\newblock
  \bibinfo{howpublished}{\url{https://www.fireeye.com/content/dam/fireeye-www/global/en/current-threats/pdfs/rpt-operation-saffron-rose.pdf}}.
    (\bibinfo{date}{{May}} \bibinfo{year}{2013}).
\newblock


\bibitem[\protect\citeauthoryear{{FireEye}}{{FireEye}}{2013b}]%
        {sunshop}
\bibfield{author}{\bibinfo{person}{{FireEye}}.}
  \bibinfo{year}{2013}\natexlab{b}.
\newblock \bibinfo{title}{{SUPPLY CHAIN ANALYSIS: From Quartermaster to
  SunshopFireEye}}.
\newblock
  \bibinfo{howpublished}{\url{https://www.fireeye.com/content/dam/fireeye-www/global/en/current-threats/pdfs/rpt-malware-supply-chain.pdf}}.
    (\bibinfo{date}{{November}} \bibinfo{year}{2013}).
\newblock


\bibitem[\protect\citeauthoryear{{FireEye}}{{FireEye}}{2014}]%
        {topwordsphishingattacks}
\bibfield{author}{\bibinfo{person}{{FireEye}}.}
  \bibinfo{year}{2014}\natexlab{}.
\newblock \bibinfo{title}{{Top Words Used in Spear Phishing Attacks }}.
\newblock   (\bibinfo{year}{2014}).
\newblock


\bibitem[\protect\citeauthoryear{{G DATA}}{{G DATA}}{2014}]%
        {too_hash}
\bibfield{author}{\bibinfo{person}{{G DATA}}.} \bibinfo{year}{2014}\natexlab{}.
\newblock \bibinfo{title}{{OPERATION ``TOOHASH'' HOW TARGETED ATTACKS WORK}}.
\newblock
  \bibinfo{howpublished}{\url{https://public.gdatasoftware.com/Presse/Publikationen/Whitepaper/EN/GDATA_TooHash_CaseStudy_102014_EN_v1.pdf}}.
    (\bibinfo{date}{{October}} \bibinfo{year}{2014}).
\newblock


\bibitem[\protect\citeauthoryear{Gabrilovich and Gontmakher}{Gabrilovich and
  Gontmakher}{2002}]%
        {Gabrilovich:2002:HA:503124.503156}
\bibfield{author}{\bibinfo{person}{Evgeniy Gabrilovich} {and}
  \bibinfo{person}{Alex Gontmakher}.} \bibinfo{year}{2002}\natexlab{}.
\newblock \showarticletitle{The homograph attack}.
\newblock \bibinfo{journal}{{\em Communucations of the ACM\/}}
  \bibinfo{volume}{45}, \bibinfo{number}{2} (\bibinfo{date}{Feb.}
  \bibinfo{year}{2002}), \bibinfo{pages}{128}.
\newblock
\showISSN{0001-0782}
\showDOI{%
\url{https://doi.org/10.1145/503124.503156}}


\bibitem[\protect\citeauthoryear{{Garera, Sujata and Provos, Niels and Chew,
  Monica and Rubin, Aviel D}}{{Garera, Sujata and Provos, Niels and Chew,
  Monica and Rubin, Aviel D}}{2007}]%
        {garera2007framework}
\bibfield{author}{\bibinfo{person}{{Garera, Sujata and Provos, Niels and Chew,
  Monica and Rubin, Aviel D}}.} \bibinfo{year}{2007}\natexlab{}.
\newblock \showarticletitle{{A framework for detection and measurement of
  phishing attacks}}. In \bibinfo{booktitle}{{\em {Proceedings of the 2007 ACM
  workshop on Recurring malcode}}}. {ACM}, \bibinfo{pages}{1--8}.
\newblock


\bibitem[\protect\citeauthoryear{Hao, Feamster, and Pandrangi}{Hao
  et~al\mbox{.}}{2011}]%
        {hao2011monitoring}
\bibfield{author}{\bibinfo{person}{Shuang Hao}, \bibinfo{person}{Nick
  Feamster}, {and} \bibinfo{person}{Ramakant Pandrangi}.}
  \bibinfo{year}{2011}\natexlab{}.
\newblock \showarticletitle{Monitoring the initial DNS behavior of malicious
  domains}. In \bibinfo{booktitle}{{\em Proceedings of the 2011 ACM SIGCOMM
  conference on Internet measurement conference}}. ACM,
  \bibinfo{pages}{269--278}.
\newblock


\bibitem[\protect\citeauthoryear{Hao, Thomas, Paxson, Feamster, Kreibich,
  Grier, and Hollenbeck}{Hao et~al\mbox{.}}{2013}]%
        {hao2013understanding}
\bibfield{author}{\bibinfo{person}{Shuang Hao}, \bibinfo{person}{Matthew
  Thomas}, \bibinfo{person}{Vern Paxson}, \bibinfo{person}{Nick Feamster},
  \bibinfo{person}{Christian Kreibich}, \bibinfo{person}{Chris Grier}, {and}
  \bibinfo{person}{Scott Hollenbeck}.} \bibinfo{year}{2013}\natexlab{}.
\newblock \showarticletitle{Understanding the domain registration behavior of
  spammers}. In \bibinfo{booktitle}{{\em Proceedings of the 2013 conference on
  Internet measurement conference}}. ACM, \bibinfo{pages}{63--76}.
\newblock


\bibitem[\protect\citeauthoryear{{Hao, Shuang and Kantchelian, Alex and Miller,
  Brad and Paxson, Vern and Feamster, Nick}}{{Hao, Shuang and Kantchelian, Alex
  and Miller, Brad and Paxson, Vern and Feamster, Nick}}{2016}]%
        {hao2016predator}
\bibfield{author}{\bibinfo{person}{{Hao, Shuang and Kantchelian, Alex and
  Miller, Brad and Paxson, Vern and Feamster, Nick}}.}
  \bibinfo{year}{2016}\natexlab{}.
\newblock \showarticletitle{{PREDATOR: Proactive Recognition and Elimination of
  Domain Abuse at Time-Of-Registration}}. In \bibinfo{booktitle}{{\em
  {Proceedings of the 2016 ACM SIGSAC Conference on Computer and Communications
  Security}}}. {ACM}, \bibinfo{pages}{1568--1579}.
\newblock


\bibitem[\protect\citeauthoryear{{Holgers, Tobias and Watson, David E. and
  Gribble, Steven D.}}{{Holgers, Tobias and Watson, David E. and Gribble,
  Steven D.}}{2006}]%
        {Holgers:2006:CTC:1267359.1267383}
\bibfield{author}{\bibinfo{person}{{Holgers, Tobias and Watson, David E. and
  Gribble, Steven D.}}} \bibinfo{year}{2006}\natexlab{}.
\newblock \showarticletitle{{Cutting through the confusion: a measurement study
  of homograph attacks}}. In \bibinfo{booktitle}{{\em {Proceedings of the 2006
  USENIX Annual Technical Conference}}}. 1.
\newblock
\showURL{%
\url{http://dl.acm.org/citation.cfm?id=1267359.1267383}}


\bibitem[\protect\citeauthoryear{{INFOSEC CONSORTIUM}}{{INFOSEC
  CONSORTIUM}}{2013}]%
        {apt_india}
\bibfield{author}{\bibinfo{person}{{INFOSEC CONSORTIUM}}.}
  \bibinfo{year}{2013}\natexlab{}.
\newblock \bibinfo{title}{{Inside Report -- APT Attacks on Indian Cyber
  Space}}.
\newblock
  \bibinfo{howpublished}{\url{http://ver007.com/tools/APTnotes/2013/Inside_Report_by_Infosec_Consortium.pdf}}.
    (\bibinfo{date}{{August}} \bibinfo{year}{2013}).
\newblock


\bibitem[\protect\citeauthoryear{{Jakobsson, Markus}}{{Jakobsson,
  Markus}}{2007}]%
        {jakobsson2007human}
\bibfield{author}{\bibinfo{person}{{Jakobsson, Markus}}.}
  \bibinfo{year}{2007}\natexlab{}.
\newblock \showarticletitle{{The human factor in phishing}}.
\newblock \bibinfo{journal}{{\em {Privacy \& Security of Consumer
  Information}\/}} \bibinfo{volume}{7}, \bibinfo{number}{1}
  (\bibinfo{year}{2007}), \bibinfo{pages}{1--19}.
\newblock


\bibitem[\protect\citeauthoryear{{Jakobsson, Markus and Tsow, Alex and Shah,
  Ankur and Blevis, Eli and Lim, Youn-Kyung}}{{Jakobsson, Markus and Tsow, Alex
  and Shah, Ankur and Blevis, Eli and Lim, Youn-Kyung}}{2007}]%
        {jakobsson2007instills}
\bibfield{author}{\bibinfo{person}{{Jakobsson, Markus and Tsow, Alex and Shah,
  Ankur and Blevis, Eli and Lim, Youn-Kyung}}.}
  \bibinfo{year}{2007}\natexlab{}.
\newblock \showarticletitle{{What instills trust? a qualitative study of
  phishing}}.
\newblock In \bibinfo{booktitle}{{\em {Financial Cryptography and Data
  Security}}}. \bibinfo{publisher}{{Springer}}, \bibinfo{pages}{356--361}.
\newblock


\bibitem[\protect\citeauthoryear{{Janos Szurdi and Balazs Kocso and Gabor Cseh
  and Jonathan Spring and Mark Felegyhazi and Chris Kanich}}{{Janos Szurdi and
  Balazs Kocso and Gabor Cseh and Jonathan Spring and Mark Felegyhazi and Chris
  Kanich}}{2014}]%
        {Szurdi:long-taile-of-typosquatting}
\bibfield{author}{\bibinfo{person}{{Janos Szurdi and Balazs Kocso and Gabor
  Cseh and Jonathan Spring and Mark Felegyhazi and Chris Kanich}}.}
  \bibinfo{year}{2014}\natexlab{}.
\newblock \showarticletitle{{The Long
  {\textquotedblleft}Taile{\textquotedblright} of Typosquatting Domain Names}}.
  In \bibinfo{booktitle}{{\em {23rd USENIX Security Symposium (USENIX Security
  14)}}}. \bibinfo{publisher}{{USENIX Association}}, \bibinfo{address}{{San
  Diego, CA}}, \bibinfo{pages}{191--206}.
\newblock
\showISBNx{978-1-931971-15-7}
\showURL{%
\url{https://www.usenix.org/conference/usenixsecurity14/technical-sessions/presentation/szurdi}}


\bibitem[\protect\citeauthoryear{{JPCERT/CC}}{{JPCERT/CC}}{2016}]%
        {asruex}
\bibfield{author}{\bibinfo{person}{{JPCERT/CC}}.}
  \bibinfo{year}{2016}\natexlab{}.
\newblock \bibinfo{title}{{Asruex: Malware Infecting through Shortcut Files}}.
\newblock
  \bibinfo{howpublished}{\url{http://blog.jpcert.or.jp/2016/06/asruex-malware-infecting-through-shortcut-files.html}}.
    (\bibinfo{date}{{June}} \bibinfo{year}{2016}).
\newblock


\bibitem[\protect\citeauthoryear{{Kaspersky}}{{Kaspersky}}{2013}]%
        {ice_frog}
\bibfield{author}{\bibinfo{person}{{Kaspersky}}.}
  \bibinfo{year}{2013}\natexlab{}.
\newblock \bibinfo{title}{{THE 'ICEFOG' APT: A TALE OF CLOAK AND THREE
  DAGGERS}}.
\newblock
  \bibinfo{howpublished}{\url{https://kasperskycontenthub.com/wp-content/uploads/sites/43/vlpdfs/icefog.pdf}}.
    (\bibinfo{date}{{September}} \bibinfo{year}{2013}).
\newblock


\bibitem[\protect\citeauthoryear{{Kaspersky}}{{Kaspersky}}{2015}]%
        {carbanak}
\bibfield{author}{\bibinfo{person}{{Kaspersky}}.}
  \bibinfo{year}{2015}\natexlab{}.
\newblock \bibinfo{title}{{CARBANAK APT THE GREAT BANK ROBBERY}}.
\newblock
  \bibinfo{howpublished}{\url{https://securelist.com/files/2015/02/Carbanak_APT_eng.pdf}}.
    (\bibinfo{date}{{February}} \bibinfo{year}{2015}).
\newblock


\bibitem[\protect\citeauthoryear{{Kaspersky Lab}}{{Kaspersky Lab}}{2014a}]%
        {darkhotel}
\bibfield{author}{\bibinfo{person}{{Kaspersky Lab}}.}
  \bibinfo{year}{2014}\natexlab{a}.
\newblock \bibinfo{title}{{DARKHOTEL INDICATORS OF COMPROMISE}}.
\newblock
  \bibinfo{howpublished}{\url{https://securelist.com/files/2014/11/darkhotelappendixindicators_kl.pdf}}.
    (\bibinfo{date}{{November}} \bibinfo{year}{2014}).
\newblock


\bibitem[\protect\citeauthoryear{{Kaspersky Lab}}{{Kaspersky Lab}}{2014b}]%
        {snake_uroboros}
\bibfield{author}{\bibinfo{person}{{Kaspersky Lab}}.}
  \bibinfo{year}{2014}\natexlab{b}.
\newblock \bibinfo{title}{{The Epic Turla Operation: Solving some of the
  mysteries of Snake/Uroboros}}.
\newblock
  \bibinfo{howpublished}{\url{https://cdn.securelist.com/files/2014/08/KL_Epic_Turla_Technical_Appendix_20140806.pdf}}.
    (\bibinfo{date}{{August}} \bibinfo{year}{2014}).
\newblock


\bibitem[\protect\citeauthoryear{{Khan, Mohammad Taha and Huo, Xiang and Li,
  Zhou and Kanich, Chris}}{{Khan, Mohammad Taha and Huo, Xiang and Li, Zhou and
  Kanich, Chris}}{2015}]%
        {khan2015every}
\bibfield{author}{\bibinfo{person}{{Khan, Mohammad Taha and Huo, Xiang and Li,
  Zhou and Kanich, Chris}}.} \bibinfo{year}{2015}\natexlab{}.
\newblock \showarticletitle{{Every Second Counts: Quantifying the Negative
  Externalities of Cybercrime via Typosquatting}}. In \bibinfo{booktitle}{{\em
  {Proceedings of the 36th IEEE Symposium on Security and Privacy}}}.
\newblock


\bibitem[\protect\citeauthoryear{{Kreibich, Christian and Kanich, Chris and
  Levchenko, Kirill and Enright, Brandon and Voelker, Geoffrey M and Paxson,
  Vern and Savage, Stefan}}{{Kreibich, Christian and Kanich, Chris and
  Levchenko, Kirill and Enright, Brandon and Voelker, Geoffrey M and Paxson,
  Vern and Savage, Stefan}}{2008}]%
        {kreibich2008spam}
\bibfield{author}{\bibinfo{person}{{Kreibich, Christian and Kanich, Chris and
  Levchenko, Kirill and Enright, Brandon and Voelker, Geoffrey M and Paxson,
  Vern and Savage, Stefan}}.} \bibinfo{year}{2008}\natexlab{}.
\newblock \showarticletitle{{On the Spam Campaign Trail.}}
\newblock \bibinfo{journal}{{\em {LEET}\/}} \bibinfo{volume}{8},
  \bibinfo{number}{2008} (\bibinfo{year}{2008}), \bibinfo{pages}{1--9}.
\newblock


\bibitem[\protect\citeauthoryear{{Let's Encrypt}}{{Let's Encrypt}}{2017}]%
        {letsencrypt}
\bibfield{author}{\bibinfo{person}{{Let's Encrypt}}.}
  \bibinfo{year}{2017}\natexlab{}.
\newblock \bibinfo{title}{{Let's Encrypt --- Free SSL/TLS Certificates}}.
\newblock \bibinfo{howpublished}{\url{https://letsencrypt.org}}.
  (\bibinfo{year}{2017}).
\newblock


\bibitem[\protect\citeauthoryear{{Lever, Chaz and Walls, Robert and Nadji,
  Yacin and Dagon, David and McDaniel, Patrick and Antonakakis, Manos}}{{Lever,
  Chaz and Walls, Robert and Nadji, Yacin and Dagon, David and McDaniel,
  Patrick and Antonakakis, Manos}}{2016}]%
        {lever2016domain}
\bibfield{author}{\bibinfo{person}{{Lever, Chaz and Walls, Robert and Nadji,
  Yacin and Dagon, David and McDaniel, Patrick and Antonakakis, Manos}}.}
  \bibinfo{year}{2016}\natexlab{}.
\newblock \showarticletitle{{Domain-Z: 28 Registrations Later}}.
\newblock  (\bibinfo{year}{2016}).
\newblock


\bibitem[\protect\citeauthoryear{{Liu, Daiping and Hao, Shuai and Wang,
  Haining}}{{Liu, Daiping and Hao, Shuai and Wang, Haining}}{2016}]%
        {liu2016all}
\bibfield{author}{\bibinfo{person}{{Liu, Daiping and Hao, Shuai and Wang,
  Haining}}.} \bibinfo{year}{2016}\natexlab{}.
\newblock \showarticletitle{{All Your DNS Records Point to Us: Understanding
  the Security Threats of Dangling DNS Records}}. In \bibinfo{booktitle}{{\em
  {Proceedings of the 2016 ACM SIGSAC Conference on Computer and Communications
  Security}}}. {ACM}, \bibinfo{pages}{1414--1425}.
\newblock


\bibitem[\protect\citeauthoryear{{Ma, Justin and Saul, Lawrence K and Savage,
  Stefan and Voelker, Geoffrey M}}{{Ma, Justin and Saul, Lawrence K and Savage,
  Stefan and Voelker, Geoffrey M}}{2009}]%
        {ma2009beyond}
\bibfield{author}{\bibinfo{person}{{Ma, Justin and Saul, Lawrence K and Savage,
  Stefan and Voelker, Geoffrey M}}.} \bibinfo{year}{2009}\natexlab{}.
\newblock \showarticletitle{{Beyond Blacklists: Learning to Detect Malicious
  Web Sites from Suspicious URLs}}. In \bibinfo{booktitle}{{\em {Proceedings of
  the 15th ACM SIGKDD International Conference on Knowledge Discovery and Data
  Mining (KDD)}}}.
\newblock


\bibitem[\protect\citeauthoryear{{Marczak, William R and Scott-Railton, John
  and Marquis-Boire, Morgan and Paxson, Vern}}{{Marczak, William R and
  Scott-Railton, John and Marquis-Boire, Morgan and Paxson, Vern}}{2014}]%
        {marczak2014governments}
\bibfield{author}{\bibinfo{person}{{Marczak, William R and Scott-Railton, John
  and Marquis-Boire, Morgan and Paxson, Vern}}.}
  \bibinfo{year}{2014}\natexlab{}.
\newblock \showarticletitle{{When governments hack opponents: A look at actors
  and technology}}. In \bibinfo{booktitle}{{\em {23rd USENIX Security Symposium
  (USENIX Security 14)}}}. \bibinfo{pages}{511--525}.
\newblock


\bibitem[\protect\citeauthoryear{{Microsoft}}{{Microsoft}}{2015}]%
        {strontium}
\bibfield{author}{\bibinfo{person}{{Microsoft}}.}
  \bibinfo{year}{2015}\natexlab{}.
\newblock \bibinfo{title}{{Microsoft Security Intelligence Report Volume 19 |
  January through June, 2015}}.
\newblock
  \bibinfo{howpublished}{\url{http://download.microsoft.com/download/4/4/C/44CDEF0E-7924-4787-A56A-16261691ACE3/Microsoft_Security_Intelligence_Report_Volume_19_English.pdf}}.
    (\bibinfo{date}{{June}} \bibinfo{year}{2015}).
\newblock


\bibitem[\protect\citeauthoryear{{Miramirkhani, Najmeh and Starov, Oleksii and
  Nikiforakis, Nick}}{{Miramirkhani, Najmeh and Starov, Oleksii and
  Nikiforakis, Nick}}{2017}]%
        {miramirkhani2017tss}
\bibfield{author}{\bibinfo{person}{{Miramirkhani, Najmeh and Starov, Oleksii
  and Nikiforakis, Nick}}.} \bibinfo{year}{2017}\natexlab{}.
\newblock \showarticletitle{{Dial One for Scam: A Large-Scale Analysis of
  Technical Support Scams}}. In \bibinfo{booktitle}{{\em {Proceedings of the
  24th Network and Distributed System Security Symposium (NDSS 2017)}}}.
  {Internet Society}.
\newblock


\bibitem[\protect\citeauthoryear{Mockapetris}{Mockapetris}{1983a}]%
        {rfc882}
\bibfield{author}{\bibinfo{person}{P.V. Mockapetris}.}
  \bibinfo{year}{1983}\natexlab{a}.
\newblock \bibinfo{title}{{Domain names: Concepts and facilities}}.
\newblock \bibinfo{howpublished}{RFC 882}.   (\bibinfo{date}{Nov.}
  \bibinfo{year}{1983}).
\newblock
\showURL{%
\url{http://www.ietf.org/rfc/rfc882.txt}}
\newblock
\shownote{Obsoleted by RFCs 1034, 1035, updated by RFC 973.}


\bibitem[\protect\citeauthoryear{Mockapetris}{Mockapetris}{1983b}]%
        {rfc883}
\bibfield{author}{\bibinfo{person}{P.V. Mockapetris}.}
  \bibinfo{year}{1983}\natexlab{b}.
\newblock \bibinfo{title}{{Domain names: Implementation specification}}.
\newblock \bibinfo{howpublished}{RFC 883}.   (\bibinfo{date}{Nov.}
  \bibinfo{year}{1983}).
\newblock
\showURL{%
\url{http://www.ietf.org/rfc/rfc883.txt}}
\newblock
\shownote{Obsoleted by RFCs 1034, 1035, updated by RFC 973.}


\bibitem[\protect\citeauthoryear{Mockapetris}{Mockapetris}{1987a}]%
        {rfc1034}
\bibfield{author}{\bibinfo{person}{P.V. Mockapetris}.}
  \bibinfo{year}{1987}\natexlab{a}.
\newblock \bibinfo{title}{{Domain names - concepts and facilities}}.
\newblock \bibinfo{howpublished}{RFC 1034 (INTERNET STANDARD)}.
  (\bibinfo{date}{Nov.} \bibinfo{year}{1987}).
\newblock
\showURL{%
\url{http://www.ietf.org/rfc/rfc1034.txt}}
\newblock
\shownote{Updated by RFCs 1101, 1183, 1348, 1876, 1982, 2065, 2181, 2308, 2535,
  4033, 4034, 4035, 4343, 4035, 4592, 5936.}


\bibitem[\protect\citeauthoryear{Mockapetris}{Mockapetris}{1987b}]%
        {rfc1035}
\bibfield{author}{\bibinfo{person}{P.V. Mockapetris}.}
  \bibinfo{year}{1987}\natexlab{b}.
\newblock \bibinfo{title}{{Domain names - implementation and specification}}.
\newblock \bibinfo{howpublished}{RFC 1035 (INTERNET STANDARD)}.
  (\bibinfo{date}{Nov.} \bibinfo{year}{1987}).
\newblock
\showURL{%
\url{http://www.ietf.org/rfc/rfc1035.txt}}


\bibitem[\protect\citeauthoryear{Moore and Edelman}{Moore and Edelman}{2010}]%
        {Moore2010}
\bibfield{author}{\bibinfo{person}{Tyler Moore} {and} \bibinfo{person}{Benjamin
  Edelman}.} \bibinfo{year}{2010}\natexlab{}.
\newblock \showarticletitle{Measuring the Perpetrators and Funders of
  Typosquatting}. In \bibinfo{booktitle}{{\em Financial Cryptography and Data
  Security}}, Vol.~\bibinfo{volume}{6052}. \bibinfo{pages}{175--191}.
\newblock


\bibitem[\protect\citeauthoryear{Nikiforakis, Acker, Meert, Desmet, Piessens,
  and Joosen}{Nikiforakis et~al\mbox{.}}{2013}]%
        {bitsquatting_www2013}
\bibfield{author}{\bibinfo{person}{Nick Nikiforakis},
  \bibinfo{person}{Steven~Van Acker}, \bibinfo{person}{Wannes Meert},
  \bibinfo{person}{Lieven Desmet}, \bibinfo{person}{Frank Piessens}, {and}
  \bibinfo{person}{Wouter Joosen}.} \bibinfo{year}{2013}\natexlab{}.
\newblock \showarticletitle{{Bitsquatting: Exploiting bit-flips for fun, or
  profit?}}. In \bibinfo{booktitle}{{\em WWW'13}}. \bibinfo{pages}{989--998}.
\newblock


\bibitem[\protect\citeauthoryear{{Nikiforakis, Nick and Balduzzi, Marco and
  Desmet, Lieven and Piessens, Frank and Joosen, Wouter}}{{Nikiforakis, Nick
  and Balduzzi, Marco and Desmet, Lieven and Piessens, Frank and Joosen,
  Wouter}}{2014}]%
        {nikiforakis2014soundsquatting}
\bibfield{author}{\bibinfo{person}{{Nikiforakis, Nick and Balduzzi, Marco and
  Desmet, Lieven and Piessens, Frank and Joosen, Wouter}}.}
  \bibinfo{year}{2014}\natexlab{}.
\newblock \showarticletitle{{Soundsquatting: Uncovering the use of homophones
  in domain squatting}}.
\newblock In \bibinfo{booktitle}{{\em {Information Security}}}.
  \bibinfo{publisher}{{Springer}}, \bibinfo{pages}{291--308}.
\newblock


\bibitem[\protect\citeauthoryear{{Nikiforakis, Nick and Van Acker, Steven and
  Meert, Wannes and Desmet, Lieven and Piessens, Frank and Joosen,
  Wouter}}{{Nikiforakis, Nick and Van Acker, Steven and Meert, Wannes and
  Desmet, Lieven and Piessens, Frank and Joosen, Wouter}}{2013}]%
        {nikiforakis2013bitsquatting}
\bibfield{author}{\bibinfo{person}{{Nikiforakis, Nick and Van Acker, Steven and
  Meert, Wannes and Desmet, Lieven and Piessens, Frank and Joosen, Wouter}}.}
  \bibinfo{year}{2013}\natexlab{}.
\newblock \showarticletitle{{Bitsquatting: Exploiting bit-flips for fun, or
  profit?}}. In \bibinfo{booktitle}{{\em {Proceedings of the 22nd international
  conference on World Wide Web}}}. {ACM}, \bibinfo{pages}{989--998}.
\newblock


\bibitem[\protect\citeauthoryear{pwc}{pwc}{2014}]%
        {scanbox}
\bibfield{author}{\bibinfo{person}{pwc}.} \bibinfo{year}{2014}\natexlab{}.
\newblock \bibinfo{title}{{ScanBox framework --- who's affected, and who's
  using it?}}
\newblock
  \bibinfo{howpublished}{\url{http://pwc.blogs.com/cyber_security_updates/2014/10/scanbox-framework-whos-affected-and-whos-using-it-1.html}}.
    (\bibinfo{date}{{October}} \bibinfo{year}{2014}).
\newblock


\bibitem[\protect\citeauthoryear{pwc}{pwc}{2015a}]%
        {poisonIvy_pwc}
\bibfield{author}{\bibinfo{person}{pwc}.} \bibinfo{year}{2015}\natexlab{a}.
\newblock \bibinfo{title}{{Attacks against Israeli \& Palestinian interests}}.
\newblock
  \bibinfo{howpublished}{\url{http://pwc.blogs.com/cyber_security_updates/2015/04/attacks-against-israeli-palestinian-interests.html}}.
    (\bibinfo{date}{{April}} \bibinfo{year}{2015}).
\newblock


\bibitem[\protect\citeauthoryear{pwc}{pwc}{2015b}]%
        {sofacy_ii}
\bibfield{author}{\bibinfo{person}{pwc}.} \bibinfo{year}{2015}\natexlab{b}.
\newblock \bibinfo{title}{{Cyber Threat Operations Sofacy II-- Same Sofacy,
  Different Day}}.
\newblock
  \bibinfo{howpublished}{\url{http://pwc.blogs.com/files/cto-tib-20150420-01a.pdf}}.
    (\bibinfo{date}{{April}} \bibinfo{year}{2015}).
\newblock


\bibitem[\protect\citeauthoryear{Rahbarinia, Perdisci, and
  Antonakakis}{Rahbarinia et~al\mbox{.}}{2015}]%
        {7266868}
\bibfield{author}{\bibinfo{person}{B. Rahbarinia}, \bibinfo{person}{R.
  Perdisci}, {and} \bibinfo{person}{M. Antonakakis}.}
  \bibinfo{year}{2015}\natexlab{}.
\newblock \showarticletitle{{Segugio: Efficient Behavior-Based Tracking of
  Malware-Control Domains in Large ISP Networks}}. In \bibinfo{booktitle}{{\em
  {Dependable Systems and Networks (DSN), 2015 45th Annual IEEE/IFIP
  International Conference on}}}. \bibinfo{pages}{403--414}.
\newblock
\showDOI{%
\url{https://doi.org/10.1109/DSN.2015.35}}


\bibitem[\protect\citeauthoryear{{root9B}}{{root9B}}{2015}]%
        {apt28}
\bibfield{author}{\bibinfo{person}{{root9B}}.} \bibinfo{year}{2015}\natexlab{}.
\newblock \bibinfo{title}{{APT28 targets Financial Markets ROOT9B RELEASES ZERO
  DAY HASHES}}.
\newblock
  \bibinfo{howpublished}{\url{https://www.root9b.com/sites/default/files/whitepapers/R9b_FSOFACY_0.pdf}}.
    (\bibinfo{date}{{May}} \bibinfo{year}{2015}).
\newblock


\bibitem[\protect\citeauthoryear{{Ryan Kelly}}{{Ryan Kelly}}{2016}]%
        {pyenchant}
\bibfield{author}{\bibinfo{person}{{Ryan Kelly}}.}
  \bibinfo{year}{2016}\natexlab{}.
\newblock \bibinfo{title}{{PyEnchant a spellchecking library for Python}}.
\newblock \bibinfo{howpublished}{\url{http://pythonhosted.org/pyenchant/}}.
  (\bibinfo{year}{2016}).
\newblock


\bibitem[\protect\citeauthoryear{{S. Krishnan and F. Monrose}}{{S. Krishnan and
  F. Monrose}}{2011}]%
        {5958207}
\bibfield{author}{\bibinfo{person}{{S. Krishnan and F. Monrose}}.}
  \bibinfo{year}{2011}\natexlab{}.
\newblock \showarticletitle{{An empirical study of the performance, security
  and privacy implications of domain name prefetching}}. In
  \bibinfo{booktitle}{{\em {Dependable Systems Networks (DSN), 2011 IEEE/IFIP
  41st International Conference on}}}. \bibinfo{pages}{61--72}.
\newblock
\showISSN{1530-0889}
\showDOI{%
\url{https://doi.org/10.1109/DSN.2011.5958207}}


\bibitem[\protect\citeauthoryear{{SecureWorks}}{{SecureWorks}}{2013}]%
        {comfoo}
\bibfield{author}{\bibinfo{person}{{SecureWorks}}.}
  \bibinfo{year}{2013}\natexlab{}.
\newblock \bibinfo{title}{{Secrets of the Comfoo Masters}}.
\newblock
  \bibinfo{howpublished}{\url{https://www.secureworks.com/research/secrets-of-the-comfoo-masters}}.
    (\bibinfo{date}{{July}} \bibinfo{year}{2013}).
\newblock


\bibitem[\protect\citeauthoryear{{Segaran, Toby and Hammerbacher,
  Jeff}}{{Segaran, Toby and Hammerbacher, Jeff}}{2009}]%
        {segaran2009beautiful}
\bibfield{author}{\bibinfo{person}{{Segaran, Toby and Hammerbacher, Jeff}}.}
  \bibinfo{year}{2009}\natexlab{}.
\newblock \bibinfo{booktitle}{{\em {Beautiful data: the stories behind elegant
  data solutions}}}.
\newblock \bibinfo{publisher}{" O'Reilly Media, Inc."}.
\newblock


\bibitem[\protect\citeauthoryear{{Snyder, Peter and Kanich, Chris}}{{Snyder,
  Peter and Kanich, Chris}}{2015}]%
        {snyder2015no}
\bibfield{author}{\bibinfo{person}{{Snyder, Peter and Kanich, Chris}}.}
  \bibinfo{year}{2015}\natexlab{}.
\newblock \showarticletitle{{No please, after you: Detecting fraud in affiliate
  marketing networks}}. In \bibinfo{booktitle}{{\em Proceedings of the Workshop
  on the Economics of Information Security (WEIS)}}.
\newblock


\bibitem[\protect\citeauthoryear{{SOWPODS}}{{SOWPODS}}{2016}]%
        {sowpods}
\bibfield{author}{\bibinfo{person}{{SOWPODS}}.}
  \bibinfo{year}{2016}\natexlab{}.
\newblock \bibinfo{title}{{SOWPODS Scrabble Word List}}.
\newblock
  \bibinfo{howpublished}{\url{https://www.wordgamedictionary.com/sowpods/}}.
  (\bibinfo{year}{2016}).
\newblock


\bibitem[\protect\citeauthoryear{{Symantec}}{{Symantec}}{2013a}]%
        {comment_crew}
\bibfield{author}{\bibinfo{person}{{Symantec}}.}
  \bibinfo{year}{2013}\natexlab{a}.
\newblock \bibinfo{title}{{Comment Crew: Indicators of Compromise}}.
\newblock
  \bibinfo{howpublished}{\url{https://www.symantec.com/content/en/us/enterprise/media/security_response/whitepapers/comment_crew_indicators_of_compromise.pdf}}.
    (\bibinfo{date}{{February}} \bibinfo{year}{2013}).
\newblock


\bibitem[\protect\citeauthoryear{{Symantec}}{{Symantec}}{2013b}]%
        {hidden_lynx}
\bibfield{author}{\bibinfo{person}{{Symantec}}.}
  \bibinfo{year}{2013}\natexlab{b}.
\newblock \bibinfo{title}{{Hidden Lynx -- Professional Hackers for Hire}}.
\newblock
  \bibinfo{howpublished}{\url{http://www.symantec.com/content/en/us/enterprise/media/security_response/whitepapers/hidden_lynx.pdf}}.
    (\bibinfo{date}{{September}} \bibinfo{year}{2013}).
\newblock


\bibitem[\protect\citeauthoryear{{Symantec}}{{Symantec}}{2016}]%
        {suckfly}
\bibfield{author}{\bibinfo{person}{{Symantec}}.}
  \bibinfo{year}{2016}\natexlab{}.
\newblock \bibinfo{title}{{A Closer Look at MiniDuke}}.
\newblock
  \bibinfo{howpublished}{\url{https://www.symantec.com/connect/blogs/indian-organizations-targeted-suckfly-attacks}}.
    (\bibinfo{date}{{May}} \bibinfo{year}{2016}).
\newblock


\bibitem[\protect\citeauthoryear{{TrendMicro}}{{TrendMicro}}{2011}]%
        {lurid}
\bibfield{author}{\bibinfo{person}{{TrendMicro}}.}
  \bibinfo{year}{2011}\natexlab{}.
\newblock \bibinfo{title}{{THE ``LURID'' DOWNLOADER}}.
\newblock
  \bibinfo{howpublished}{\url{http://la.trendmicro.com/media/misc/lurid-downloader-enfal-report-en.pdf}}.
    (\bibinfo{date}{{September}} \bibinfo{year}{2011}).
\newblock


\bibitem[\protect\citeauthoryear{{TrendMicro}}{{TrendMicro}}{2014}]%
        {evilgrab}
\bibfield{author}{\bibinfo{person}{{TrendMicro}}.}
  \bibinfo{year}{2014}\natexlab{}.
\newblock \bibinfo{title}{{2Q Report on Targeted Attack Campaigns }}.
\newblock
  \bibinfo{howpublished}{\url{http://la.trendmicro.com/media/misc/lurid-downloader-enfal-report-en.pdf}}.
    (\bibinfo{date}{{January}} \bibinfo{year}{2014}).
\newblock


\bibitem[\protect\citeauthoryear{{TrendMicro}}{{TrendMicro}}{2016a}]%
        {pawn_storm}
\bibfield{author}{\bibinfo{person}{{TrendMicro}}.}
  \bibinfo{year}{2016}\natexlab{a}.
\newblock \bibinfo{title}{{Looking Into a Cyber-Attack Facilitator in the
  Netherlands }}.
\newblock
  \bibinfo{howpublished}{\url{http://documents.trendmicro.com/assets/appendix_looking-into-a-cyber-attack-facilitator-in-the-netherlands.pdf}}.
    (\bibinfo{date}{April} \bibinfo{year}{2016}).
\newblock


\bibitem[\protect\citeauthoryear{{TrendMicro}}{{TrendMicro}}{2016b}]%
        {trendmicrocheck}
\bibfield{author}{\bibinfo{person}{{TrendMicro}}.}
  \bibinfo{year}{2016}\natexlab{b}.
\newblock \bibinfo{title}{{Securing Your Journey to the Cloud}}.
\newblock \bibinfo{howpublished}{\url{http://www.trendmicro.com/}}.
  (\bibinfo{year}{2016}).
\newblock


\bibitem[\protect\citeauthoryear{Vissers, Joosen, and Nikiforakis}{Vissers
  et~al\mbox{.}}{2015}]%
        {parking_ndss2015}
\bibfield{author}{\bibinfo{person}{Thomas Vissers}, \bibinfo{person}{Wouter
  Joosen}, {and} \bibinfo{person}{Nick Nikiforakis}.}
  \bibinfo{year}{2015}\natexlab{}.
\newblock \showarticletitle{{Parking Sensors: Analyzing and Detecting Parked
  Domains}}. In \bibinfo{booktitle}{{\em Proceedings of the 22nd Network and
  Distributed System Security Symposium (NDSS)}}.
\newblock


\bibitem[\protect\citeauthoryear{{Wang, Yi-Min and Beck, Doug and Wang, Jeffrey
  and Verbowski, Chad and Daniels, Brad}}{{Wang, Yi-Min and Beck, Doug and
  Wang, Jeffrey and Verbowski, Chad and Daniels, Brad}}{2006}]%
        {wang2006strider}
\bibfield{author}{\bibinfo{person}{{Wang, Yi-Min and Beck, Doug and Wang,
  Jeffrey and Verbowski, Chad and Daniels, Brad}}.}
  \bibinfo{year}{2006}\natexlab{}.
\newblock \showarticletitle{{Strider Typo-Patrol: Discovery and Analysis of
  Systematic Typo-Squatting.}}
\newblock \bibinfo{journal}{{\em {SRUTI}\/}}  \bibinfo{volume}{6}
  (\bibinfo{year}{2006}), \bibinfo{pages}{31--36}.
\newblock


\bibitem[\protect\citeauthoryear{Weimer}{Weimer}{2005}]%
        {weimer05}
\bibfield{author}{\bibinfo{person}{F. Weimer}.}
  \bibinfo{year}{2005}\natexlab{}.
\newblock \showarticletitle{Passive {DNS} Replication}. In
  \bibinfo{booktitle}{{\em Proceedings of FIRST Conference on Computer Security
  Incident}}. \bibinfo{address}{Hand ling, Singapore}.
\newblock


\bibitem[\protect\citeauthoryear{{Y. Chen and M. Antonakakis and R. Perdisci
  and Y. Nadji and D. Dagon and W. Lee}}{{Y. Chen and M. Antonakakis and R.
  Perdisci and Y. Nadji and D. Dagon and W. Lee}}{2014}]%
        {6903614}
\bibfield{author}{\bibinfo{person}{{Y. Chen and M. Antonakakis and R. Perdisci
  and Y. Nadji and D. Dagon and W. Lee}}.} \bibinfo{year}{2014}\natexlab{}.
\newblock \showarticletitle{{DNS Noise: Measuring the Pervasiveness of
  Disposable Domains in Modern DNS Traffic}}. In \bibinfo{booktitle}{{\em
  {Dependable Systems and Networks (DSN), 2014 44th Annual IEEE/IFIP
  International Conference on}}}. \bibinfo{pages}{598--609}.
\newblock
\showDOI{%
\url{https://doi.org/10.1109/DSN.2014.61}}


\end{thebibliography}
\break
\appendix
\section{Appendix}
\label{section:appendix}

\subsection{Selected Trademarks}

Table~\ref{tab:categories} depicts the categories and respective number of
trademarks for each category we used to identify combosquatting domain names.
The second column provides an example of a trademark for each category.

\begin{table}[b] \centering
  \begin{tabular}{l l l r}
    \textbf{Category} & \textbf{Example} & \multicolumn{1}{c}{\textbf{Count}} \\
    \hline
    Adult Content & youporn[.]com & 11 \\
    Blogging & blogspot[.]com & 22 \\
    Computers & adobe[.]com & 10 \\
    Couriers & fedex[.]com & 1 \\
    E-Learning & wikipedia[.]org & 12 \\
    E-Shop (Auctions) & craigslist[.]org & 3 \\
    E-Shop (Online) & amazon[.]com & 16 \\
    E-Shop (Physical) & costco[.]com & 21 \\
    Energy & chevron[.]com & 15 \\
    File Sharing & dropbox[.]com & 4 \\
    Financial & paypal[.]com & 17 \\
    Lifestyle & imdb[.]com & 19 \\
    News & nytimes[.]com & 32 \\
    Photography & tumblr[.]com & 9 \\
    Politics & democraticunderground[.]com & 7 \\
    Radio \& TV & netflix[.]com & 4 \\
    Search Engines & google[.]com & 6 \\
    Social Networks & facebook[.]com & 7 \\
    Software \& Web & office365[.]com & 34 \\
    Streaming & youtube[.]com & 7 \\
    Telecom & comcast[.]net & 4 \\
    Travel & expedia[.]com & 7 \\
    \hline
  \end{tabular}
  \caption{Trademark business categories.}
  \label{tab:categories}
\end{table}

\subsection{Most Frequent Words per Category}

Table~\ref{table:words} summarizes the ten most frequent words for each
trademark category. There, we see that many of the popular words closely
correlate with the type of trademark being abused, such as, the words
\textit{apple, game,} and \textit{phones} being popular in the
``Computers/Internet'' category and the words \textit{president, vote,} and
\textit{elect} in the ``Politics'' category.

\begin{table*}[t]
	\begin{center}
    \scalebox{0.8}{
      \begin{tabular}{lllllllllll}
        \textbf{Category} & \multicolumn{10}{c}{\textbf{Most Frequent Words}} \\
        \hline
        \textbf{Adult Content}   &free &xxx &porn &sex &gay &live &tube &porno &videos &hot \\ 
        \textbf{Blogging}   & fuck  & yeah & love  & themes   &free &theme &life &blog &best &just \\
        \textbf{Computers}  & apple &games &phones &galaxy &phone &office &free &online &support &home \\
        \textbf{Couriers}    &office &ground &online &freight &delivery &express &shipping &print &services &service \\
        \textbf{E-Learning}    &club &square &school &business &university &health &group &property &online &pilgrim \\
        \textbf{E-Shop (Auctions)} &cars &car &sale &account &south &new &post &posting &san &jobs \\
        \textbf{E-Shop (Online)}    &line &store &kindle &online &shop &free &deals &best &lay &card \\
        \textbf{E-Shop (Physical)} &price &sale &store &card &online &prices &home &stores &shop &cheap \\
        \textbf{Energy}  &card &cards &online &business &tex &credit &energy &account &chemical &gift \\
        \textbf{File Sharing} &movie &movies &file &free &archive &user &content &login &online &watch \\
        \textbf{Financial}   &bank &online &investment &service &account &services &card &worldwide &mortgage &update\\
        \textbf{Lifestyle}   &world &land &channel &vacation &games &princess &movie &villa &paris &club\\  
        \textbf{News}   &news &mike &online &zine &foundation &com &family &new &trust &media\\ 
        \textbf{Photography}   &marketing &photography &photo &buy &time &followers &family &com &photos &best\\
        \textbf{Politics} & president & vote & elect & official & campaign & trump & truth & com & stop & sucks \\
        \textbf{Radio \& TV}   &free &movies &watch &xxx &movie &chill &account &login &canada &new\\  
        \textbf{Search Engines} &plus &mail &search &glass &free &apps &com &play &maps &google\\
        \textbf{Social Networks}   &marketing &followers &free &login &buy &account &page &com &business &apps\\
        \textbf{Software \& Web}  &best &county &new &online &mobile &home &free &sucks &beach &city\\
        \textbf{Streaming}   &video &videos &free &download &music & views &converter &best &buy &listen\\
        \textbf{Telecom}  &wireless &universal &phone &business &wire &center &online &phones &free &net\\
        \textbf{Travel}   &head &island &paris &hotel &garden &inn &hotels &estate &real &beach\\
        \hline
      \end{tabular}
    }
    \caption{Most frequent words per trademark category.}
    \label{table:words}
  \end{center}
\end{table*}

Moreover, we want to highlight the lists of words in the \textit{Couriers} and
\textit{Financial} categories. Several of the trademarks used in both categories
have been victims of spear phishing attacks according to Garera et
al.~\cite{garera2007framework}. Additionally, words like \textit{tracking,
  delivery, service,} and \textit{account} are used in both the creation of
phishing domains and phishing emails ~\cite{topwordsphishingattacks}. These
results clearly indicate that most registered combosquatting domains have been
carefully constructed by attackers to match the expected context of each abused
trademark and can be used for a variety of purposes, ranging from trademark
abuse to phishing and spear-phishing campaigns.

\subsection{Combosquatting APT Domains}

Table~\ref{tab:apt-domains} shows a list of combosquatting domain names related
to Advanced Persistent Threats (APT). These domains were found in the public APT
reports available at \url{http://tinyurl.com/apt-reports} and our $CP$ and $CA$
datasets (Table~\ref{table:DS}).

\begin{table*}[b]
	\begin{center}
    \scalebox{0.84}{
      \begin{tabular}{llllll}
        \multicolumn{1}{c}{\textbf{Trademark}} & \multicolumn{1}{c}{\textbf{Domain}} & \multicolumn{1}{c}{\textbf{APT}}& \multicolumn{1}{c}{\textbf{Activity Period}} &\multicolumn{1}{c}{\textbf{Attribution}} &\multicolumn{1}{c}{\textbf{Reference}}\\
        \hline 
        Adobe &adobearm[.]com &DarkHotel &5/12 - 11/14 &Unknown Actor & ~\cite{darkhotel}\\
        Adobe &adobekr[.]com  &Dust Storm &5/10 - 2/16 &Unknown Actor &~\cite{dust_storm} \\
        Adobe &adobeplugs[.]net &DarkHotel &5/12 - 11/14 &Unknown Actor &~\cite{darkhotel}\\
        Adobe &adobeservice[.]net &TooHash &Unknown - 10/14&Chinese Origin &~\cite{too_hash} \\
        Adobe &adobeupdates[.]com &DarkHotel &5/12 - 11/14 &Unknown Actor &~\cite{darkhotel}\\
        Adobe &adobeus[.]com &Dust Storm &5/10 - 2/16 &Unknown Actor &~\cite{dust_storm}\\
        Adobe &plugin-adobe[.]com &Saffron Rose &Unknown - 5/14 &Iranian Origin &~\cite{saffron_rose}\\
        Amazon &amazonwikis[.]com &Dust Storm &5/10 - 2/16 &Unknown Actor &~\cite{dust_storm}\\
        Delta &deltae[.]com[.]br &Comment Crew  &Unknown - 2/13&Unknown Actor &~\cite{comment_crew}\\
        Delta &deltateam[.]ir &Snake/Uroboros &Uknown - 8/14 &Unknown Actor &~\cite{snake_uroboros} \\
        Delta &leveldelta[.]com &MiniDuke &2/13 - 5/13 &Unknown Actor &~\cite{miniduke}\\
        Dropbox &online-dropbox[.]com &Asruex  &10/15 - 6/16 &Unknown Actor &~\cite{asruex}\\
        Facebook &privacy-facebook[.]me &Pawn Storm &2/16 - 4/16	&Unknown Actor &~\cite{pawn_storm}\\
        Facebook &users-facebook[.]com &Saffron Rose &Unknown - 5/14 &Iranian Origin &~\cite{saffron_rose}\\
        Facebook &xn-{}-facebook-06k[.]com &Saffron Rose &Unknown - 5/14 &Iranian Origin &~\cite{saffron_rose}\\
        Google &all-google[.]com &SpyNet &Unknown - 8/14 &Unknown Actor &~\cite{marczak2014governments}\\
        Google &drive-google[.]co &Rocket Kitten &Unknown - 11/15 &Iranian Origin &~\cite{rocket_kitten} \\
        Google &drives-google[.]co &Rocket Kitten &Unknown - 11/15 &Iranian Origin &~\cite{rocket_kitten}\\
        Google &google-blogspot[.]com &Quartermaster/Sunshop  &5/13 - 11/13 &Chinese Origin &~\cite{sunshop}\\
        Google &google-config[.]com &Comfoo &Unknown - 7/13 &Unknown Actor &~\cite{comfoo} \\
        Google &google-dash[.]com &Turbo Twist  &4/16 - 4/16 &C0d0s0 Team  &~\cite{turbo_twist}\\
        Google &google-login[.]com &Comfoo &Unknown - 7/13 &Unknown Actor &~\cite{comfoo}\\
        Google &google-office[.]com &Enfal &Unknown - 9/11 &Chinese Origin &~\cite{lurid}\\
        Google &google-officeonline[.]com &Enfal &Unknown - 9/11 &Chinese Origin &~\cite{lurid}\\
        Google &google-setting[.]com &Rocket Kitten &Unknown - 11/15 &Iranian Origin &~\cite{rocket_kitten}\\
        Google &google-verify[.]com &Rocket Kitten &Unknown - 11/15 &Iranian Origin &~\cite{rocket_kitten}\\
        Google &googlecaches[.]com &ScanBox &9/14 - 10/14 &Unknown Actor &~\cite{scanbox}\\
        Google &googlenewsup[.]net &Roaming Tiger &Unknown - 7/14 &Chinese Origin &~\cite{roaming_tiger}\\
        Google &googlesale[.]net &Ixeshe &3/14 - 6/14&Chinese Origin &~\cite{ixeshe1} \\
        Google &googlesetting[.]com &Sofacy &4/15 - 5/15 &Russian Origin &~\cite{apt28}\\
        Google &googletranslatione[.]com &Trochilus &6/15 - 1/16 &Unknown Actor &~\cite{trochilus}\\
        Google &googleupdate[.]hk &Comfoo &Unknown - 7/13 &Unknown Actor &~\cite{comfoo}\\
        Google &googlewebcache[.]com &ScanBox &9/14 - 10/14 &Unknown Actor &~\cite{scanbox}\\
        Google &imggoogle[.]com  &DarkHotel &5/22 - 11/14 &Unknown Actor &~\cite{darkhotel}\\
        Google &privacy-google[.]com &Saffron Rose &Unknown - 5/14 &Iranian Origin &~\cite{saffron_rose}\\
        Google &webmailgoogle[.]com &ScanBox &9/14 - 10/14 &Unknown Actor &~\cite{scanbox}\\
        Google &xn-{}-google-yri[.]com &Saffron Rose &Unknown - 5/14 &Iranian Origin &~\cite{saffron_rose}\\
        iCloud &localiser-icloud[.]com &Pawn Storm &2/16 - 4/16	&Unknown Actor &~\cite{pawn_storm}\\
        iCloud &securityicloudservice[.]com &Pawn Storm &2/16 - 4/16	&Unknown Actor &~\cite{pawn_storm}\\
        Microsoft &ftpmicrosoft[.]com &Quartermaster/Sunshop  &5/13 - 11/13 &Chinese Origin &~\cite{sunshop}\\
        Microsoft &microsoft-cache[.]com &Turbo Twist  &4/16 - 4/16 &C0d0s0 Team  &~\cite{turbo_twist}\\
        Microsoft &microsoft-security-center[.]com &Suckfly &7/15 - 5/16 &Unknown Actor &~\cite{suckfly}\\
        Microsoft &microsoft-xpupdate[.]com &DarkHotel &5/12 - 11/14 &Unknown Actor &~\cite{darkhotel}\\
        Microsoft &microsoftc1pol361[.]com &Carbanak &10/14 - 2/15 &Unknown Actor &~\cite{carbanak}\\
        Microsoft &microsoftmse[.]com &Four Element Sword &10/14 - 4/16 &Unknown Actor &~\cite{four_element_sword}\\
        Mozilla &mozillacdn[.]com &Poseidon Group &Unknown - 2/16 &Poseidon Group &~\cite{four_element_sword}\\
        Reuters &reuters-press[.]com &Pawn Storm &2/16 - 4/16	&Unknown Actor &~\cite{pawn_storm}\\
        Skype &downloadskype[.]cf &PoisonIvy &6/14 - 4/15 &Israelian Origin &~\cite{poisonIvy_pwc}\\
        Yahoo &cc-yahoo-inc[.]org &Pawn Storm &2/16 - 4/16	&Unknown Actor &~\cite{pawn_storm}\\
        Yahoo &delivery-yahoo[.]com &Sofacy II &Unknown - 4/15 &Unknown Actor &~\cite{sofacy_ii} \\
        Yahoo &edit-mail-yahoo[.]com &Pawn Storm &2/16 - 4/16	&Unknown Actor &~\cite{pawn_storm}\\
        Yahoo &help-yahoo-service[.]com &Pawn Storm &2/16 - 4/16	&Unknown Actor &~\cite{pawn_storm}\\
        Yahoo &newesyahoo[.]com &Apt Against India &Unknown - 8/13 &Unknown Actor &~\cite{apt_india}\\
        Yahoo &privacy-yahoo[.]com &Sofacy II &Unknown - 4/15 &Unknown Actor &~\cite{sofacy_ii}\\
        Yahoo &settings-yahoo[.]com &Sofacy II &Unknown - 4/15 &Unknown Actor &~\cite{sofacy_ii}\\
        Yahoo &us-mg6mailyahoo[.]com &Strontium &Unknown - 11/15 &Unknown Actor &~\cite{strontium}\\
        Yahoo &yahoo-config[.]com &Comfoo &Unknown - 7/13 &Unknown Actor &~\cite{comfoo}\\
        Yahoo &yahoo-user[.]com &Comfoo &Unknown - 7/13 &Unknown Actor &~\cite{comfoo}\\
        Yahoo &yahooeast[.]net &Hidden Lynx &Unknown - 9/13 &Unknown Actor &~\cite{hidden_lynx}\\
        Yahoo &yahooip[.]net  &EvilGrab  &9/13 - 1/14 &Unknown Actor &~\cite{evilgrab}\\
        Yahoo &yahoomail[.]com[.]co &Saffron Rose &Unknown - 5/14 &Iranian Origin &~\cite{saffron_rose}\\
        Yahoo &yahooprotect[.]com &EvilGrab  &9/13 - 1/14 &Unknown Actor &~\cite{evilgrab}\\
        Yahoo &yahooprotect[.]net &EvilGrab  &9/13 - 1/14 &Unknown Actor &~\cite{evilgrab}\\
        Yahoo &yahooservice[.]biz &DarkHotel &5/12 - 11/14 &Unknown Actor &~\cite{darkhotel}\\
        Yahoo &yahoowebnews[.]com &IceFrog &Unknown - 9/13 &Chinese Origin &~\cite{ice_frog} \\
        \hline
      \end{tabular}
    }
    \caption{Combosquatting domains related to APT.}
    \label{tab:apt-domains}
	\end{center}
\end{table*}

\end{document}